\documentclass[fleqn,usenatbib]{mnras}

\usepackage{amssymb}	% Extra maths symbols
\usepackage{newtxtext,newtxmath}
\usepackage{amsfonts}
\usepackage[T1]{fontenc}
\usepackage{ae,aecompl}
\usepackage[position=top]{subfig}
\usepackage{float}
\usepackage{comment}
\usepackage[normalem]{ulem}
\usepackage{bm}

\usepackage{booktabs}
\usepackage[position=top]{subfig}
\usepackage{grffile}% A pacakge which changes the algorithm to check for known extensions, so that we can insert pdf figures into this text.
\usepackage{etoolbox}
\makeatletter % Workaround to only color the year for cite commands
  \patchcmd{\NAT@citex}
    {\@citea\NAT@hyper@{%
      \NAT@nmfmt{\NAT@nm}%
      \hyper@natlinkbreak{\NAT@aysep\NAT@spacechar}{\@citeb\@extra@b@citeb}%
      \NAT@date}}
    {\@citea\NAT@nmfmt{\NAT@nm}%
    \NAT@aysep\NAT@spacechar\NAT@hyper@{\NAT@date}}{}{}

  \patchcmd{\NAT@citex}
    {\@citea\NAT@hyper@{%
      \NAT@nmfmt{\NAT@nm}%
      \hyper@natlinkbreak{\NAT@spacechar\NAT@@open\if*#1*\else#1\NAT@spacechar\fi}%
        {\@citeb\@extra@b@citeb}%
      \NAT@date}}
    {\@citea\NAT@nmfmt{\NAT@nm}%
    \NAT@spacechar\NAT@@open\if*#1*\else#1\NAT@spacechar\fi\NAT@hyper@{\NAT@date}}
    {}{}
\makeatother

\defcitealias{Brown2019}{B19}

 \makeatother
\usepackage{graphicx}	% Including figure files
\usepackage{amsmath}	% Advanced maths commands
\usepackage{pifont}
\usepackage{xcolor}
\hypersetup{allcolors=teal}
\newcommand{\sacode}{\textsc{tlusty}\ }

\title[Impact of Pop III CHE on 21-cm signal and EoR]{Effects of chemically homogeneous evolution of the first stars on the 21-cm signal and reionization}

\author[B. Liu et al.]{Boyuan Liu\textsuperscript{\href{https://orcid.org/0000-0002-4966-7450}{\includegraphics[width=2.5mm]{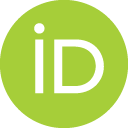}}\,}\thanks{E-mail: boyuan.liu@uni-heidelberg.de}$^{1,2}$, 
Daniel Kessler\textsuperscript{\href{https://orcid.org/0009-0006-5879-2727}{\includegraphics[width=2.5mm]{orcid.png}}\,}$^{3,4}$, 
Thomas Gessey-Jones\textsuperscript{\href{https://orcid.org/0000-0003-2138-5192}{\includegraphics[width=2.5mm]{orcid.png}}\,}$^{3,5}$, 
Jiten Dhandha\textsuperscript{\href{https://orcid.org/0000-0002-1481-0907}{\includegraphics[width=2.5mm]{orcid.png}}\,}\thanks{E-mail: jvd29@cam.ac.uk}$^{1,5}$, 
\newauthor
Anastasia Fialkov\textsuperscript{\href{https://orcid.org/0000-0002-1369-633X}{\includegraphics[width=2.5mm]{orcid.png}}\,}\thanks{E-mail: afialkov@ast.cam.ac.uk}$^{1,5}$, 
Yves Sibony\textsuperscript{\href{https://orcid.org/0000-0003-0378-4843}{\includegraphics[width=2.5mm]{orcid.png}}\,}$^{6}$, 
Georges Meynet\textsuperscript{\href{https://orcid.org/0000-0001-6181-1323}{\includegraphics[width=2.5mm]{orcid.png}}\,}$^{6}$, Volker Bromm\textsuperscript{\href{https://orcid.org/0000-0003-0212-2979}{\includegraphics[width=2.5mm]{orcid.png}}\,}$^{7,8}$, and Rennan Barkana\textsuperscript{\href{https://orcid.org/0000-0002-1557-693X}{\includegraphics[width=2.5mm]{orcid.png}}\,}$^{9}$
\\
$^{1}$Institute of Astronomy, University of Cambridge, Madingley Road, Cambridge, CB3 0HA, UK\\
$^{2}$Institut für Theoretische Astrophysik, Zentrum für Astronomie, Universität Heidelberg, Albert Ueberle Straße 2, D-69120 Heidelberg, Germany\\
$^{3}$Astrophysics Group, Cavendish Laboratory, J. J. Thomson Avenue, Cambridge, CB3 0HE, UK\\
$^{4}$School of Mathematical and Physical Sciences, University of Sheffield, Hounsfield Road, Sheffield, S3 7RH, UK\\
$^{5}$Kavli Institute for Cosmology, Madingley Road, Cambridge, CB3 0HA, UK\\
$^{6}$Observatoire de Genève, Chemin Pegasi 51, 1290 Versoix, Switzerland\\
$^{7}$Department of Astronomy, University of Texas, Austin, TX 78712, USA\\
$^{8}$Weinberg Institute for Theoretical Physics, University of Texas, Austin, TX 78712, USA\\
$^{9}$Department of Astrophysics, School of Physics and Astronomy, Tel Aviv University, Tel Aviv 69978, Israel
}
\date{Accepted XXX. Received YYY; in original form ZZZ}

\pubyear{2025}

\begin{document}
\label{firstpage}
\pagerange{\pageref{firstpage}--\pageref{lastpage}}
\maketitle

\begin{abstract} % 246 words
The first generation of stars, known as Population III (Pop~III), played a crucial role in the early Universe through their unique formation environment and metal-free composition. These stars can undergo chemically homogeneous evolution (CHE) due to fast rotation, becoming more compact and hotter/bluer than their (commonly assumed) non-rotating counterparts. In this study, we investigate the impact of Pop~III CHE on the 21-cm signal and cosmic reionization under various assumptions on Pop~III star formation, such as their formation efficiency, initial mass function, and transition to metal-enriched star formation. We combine stellar spectra computed by detailed atmosphere models with semi-numerical simulations of Cosmic Dawn and the Epoch of Reionization ($z\sim 6-30$). The key effect of CHE arises from the boosted ionizing power of Pop~III stars, which reduces the Pop III stellar mass density required to reproduce the observed Thomson scattering optical depth by a factor of $\sim 2$. Meanwhile, the maximum 21-cm global absorption signal is shallower by up to $\sim 15$ mK (11\%), partly due to the reduced Lyman-band emission from CHE, and the large-scale ($k\sim 0.2\ \rm cMpc^{-1}$) power drops by a factor of a few at $z\gtrsim 25$. In general, the effects of CHE can be comparable to those of Pop~III star formation parameters, showing an interesting interplay with distinct features in different epochs. These results highlight the importance of metal-free/poor stellar evolution in understanding the early Universe and suggest that future studies should consider joint constraints on the physics of star/galaxy formation and stellar evolution.
\end{abstract} % 246 words


%%%%%%%%%%%%%%%%%%%%%%%%%%%%%%%%%%%%%%%%%%%%%%%%%%

%%%%%%%%%%%%%%%%% BODY OF PAPER %%%%%%%%%%%%%%%%%%

\section{Introduction}\label{sec:intro}

The first generation of stars, the so-called Population~III (Pop~III), are expected to be fundamentally different from their present-day Population~I/II (Pop~I/II) counterparts due to their unique primordial formation environment and metal-free nature \citep[reviewed by, e.g.,][]{Bromm2009,Bromm2013,Haemmerle2020,Klessen2023}. As they form in primordial gas with inefficient cooling, Pop~III stars tend to be more massive than Pop~I/II stars. Their initial mass function (IMF) is typically broad and top-heavy, extending to $\sim 10^{3}\ \rm M_{\odot}$ and even higher masses ($\sim 10^{4}-10^{6}\ \rm M_{\odot}$) in extreme cases \citep[e.g.,][]{Greif2011,Greif2012,Susa2014,Hirano2014,Hirano2015,Hirano2018sc,Stacy2016,Hirano2017,Susa2019,Sugimura2020,Wollenberg2020,Chon2021,Sharda2020,Sharda2021,Sharda2022,Latif2022,Riaz2022sf,Prole2022res,Prole2023,Toyouchi2023}. Besides, if magnetic braking is inefficient\footnote{The roles played by magnetic fields in Pop~III star formation including their impact on fragmentation of primordial star-forming disks and initial spins of Pop~III stars are still in debate \citep{McKee2020,Sharda2020,Sharda2021,Sharda2022,Hirano2021,Prole2022,Saad2022,Stacy2022,Hirano2022,Sadanari2023,Sadanari2024,Sharda2024}. In extreme cases, the rotation of Pop~III (proto)stars can be slowed down by exponentially amplified magnetic fields \citep{Hirano2022}.} \citep[][]{Stacy2011,Stacy2013,Hirano2018,Kimura2023}, Pop~III stars will be born as fast rotators due to rapid accretion ($\sim 0.01-1\ \rm M_\odot\ yr^{-1}$) of gas with high angular momentum from hot, thick star-forming disks in the protostar phase. The initial rotation velocities can reach a fraction of $\sim 0.5-1$ of the critical value (i.e., equatorial Keplerian velocity).

Under such peculiar conditions and due to the lack of metals, stellar evolution and feedback of Pop~III stars are also very different from those of Pop~I/II stars \citep[e.g.,][]{Schaerer2002,Meynet2006,Ekstrom2008,Heger2010,Yoon2012,Tanikawa2020,Murphy2021,Aryan2023,Martinet2023,Nandal2023,Volpato2023,Costa2025}, producing unique signatures in direct observations \citep[e.g., strong HeII and Lyman-$\alpha$ emission and extremely blue UV spectra,][]{Windhorst2018,Grisdale2021,Nakajima2022,Vikaeus2022,Trussler2023,Katz2023,Larkin2023,Venditti2024,Zackrisson2011,Zackrisson2012,Zackrisson2023,Lecroq2025}\footnote{Although several promising candidates for Pop~III systems have been detected \citep[e.g.,][]{Welch2022,Schauer2022,Wang2022jwst,Vanzella2023,Maiolino2023,Zackrisson2023,Fujimoto2025}, direct observations of a representative sample of Pop~III stars are still challenging in the near future \citep{Gardner2006,Angel2008,Rhodes2020,Schauer2020,Katz2023,Nakajima2022,Riaz2022}.} and imprints in cosmic chemical and thermal evolution (see, e.g., \citealt{Karlsson2013,Frebel2015,Barkana2016,Dayal2018} for reviews), which allow us to constrain their properties through indirect probes. 

For instance, the chemical imprints of metal enrichment from Pop~III stars can be recorded in extremely metal-poor ($\rm [Fe/H]\lesssim -3$) stars in the local Universe as bona-fide second-generation stars, which provide essential hints on the mass distribution, nucleosynthesis, and supernova (SN) properties of Pop~III stars {\citep[e.g.,][]{Frebel2015,Ji2015,Ishigaki2018,Rossi2021,Rossi2023,Rossi2024,Rossi2024cno,Koutsouridou2023,Vanni2023}}. In fact, nucleosynthesis features of strong mixing induced by fast rotation in massive stars are found in the chemical patterns of extremely metal-poor stars observed in the local Universe \citep[e.g.,][]{Chiappini2006,Chiappini2011,Chiappini2013,Maeder2015,Choplin2017,Choplin2019,Liu2021wind,Jeena2023}, which supports the massive and fast-rotating nature of Pop~III stars. The strong nitrogen and carbon enhancement in high-$z$ galaxies observed by the James Webb Space Telescope \citep[JWST,][]{Bunker2023,D'Eugenio2023,Cameron2023,Senchyna2023,Ji2024,Schaerer2024,Sodini2024,Topping2024} can also be explained by the peculiar metal yields of fast-rotating Pop~III stars \citep{Nandal2024,Tsiatsiou2024}. 
In particular, the very high carbon (C) enhancement in the most metal-poor ($\rm [Fe/H]\lesssim -4$) stars \citep[see, e.g.,][]{Yoon2016,Yoon2018,Hansen2019,Dietz2021,Zepeda2022} can be explained by the carbon-rich (post main sequence) winds from fast-rotating massive Pop~III stars \citep{Liu2021wind,Jeena2023}. Moreover, the highest C enhancement seen in observations with absolute abundances\footnote{$A({\rm C})\equiv \log(N_{\rm C}/N_{\rm H})+12$, where $N_{\rm C}$ and $N_{\rm H}$ are the number abundances of carbon and hydrogen, respectively. Throughout this paper, $\log$ denotes the logarithm of base 10.} $A({\rm C})\gtrsim 7$ requires Pop~III stars to reach a chemically homogeneous (CH) state that can significantly boost carbon production \citep{Jeena2023}. 
%The
 %\footnote{The strong nitrogen and carbon enrichment in high-$z$ galaxies observed by the James Webb Space Telescope \citep{D'Eugenio2023,Cameron2023,Senchyna2023,Ji2024,Schaerer2024,Topping2024} can also be explained by the peculiar metal yields of fast-rotating Pop~III stars \citep{Tsiatsiou2024}.}
%, which can be common given the high initial rotation rates and weak angular momentum loss from stellar winds of Pop~III stars. \color{orange}

Such chemically-homogeneous evolution (CHE) is the extreme limit of efficient mixing from rotational-induced instabilities \citep[e.g.,][]{Yoon2006,Brott2011,Szecsi2015,Szecsi2022}, which has interesting consequences on the properties, feedback, and observational signatures of massive stars \citep{Eldridge2012,Szecsi2015,Szecsi2022,Szecsi2025,Kubatova2019,Sibony2022,Liu2024che} and their remnants, such as Wolf-Rayet stars \citep[]{Martins2009,Martins2013,Boco2025}, pair-instability supernovae \citep{Yoon2012,duBuisson2020,Umeda2024}, gamma-ray bursts \citep{Yoon2005,Yoon2006,Yoon2012}, and binary compact object mergers \citep{deMink2016,Mandel2016,Marchant2017,Marchant2023,duBuisson2020,Riley2021,Qin2023,Vigna-Gomez2025}. %CHE can be common for massive Pop~III stars with high initial rotation rates as well as weak mass and angular momentum loss from line-driven winds. 
Beyond unique metal enrichment, Pop~III stars with CHE will be more compact and hotter than in the (standard) non-rotating (NR) case and can potentially burn most of their hydrogen into helium during their main sequence (MS), so that their UV emission will also be different. It is shown in \citet{Sibony2022} that CHE significantly boosts the emission of ionizing photons from Pop~III stars (and their escape from minihaloes), which accelerates cosmic reionization, producing a Thomson scattering optical depth up to $5\sigma$ higher than the observed value $\tau_{0}=0.0544\pm0.0073$ \citep{PlanckCollaboration2020} in the most extreme case of high star formation efficiency (SFE)\footnote{Strong emission of ionizing photons from Pop~III stars can result in a double-reionization scenario \citep{Cen2003} with the first full ionization event happening at $z\gtrsim 10$ \citep{Salvador-Sole2017}, which can explain recent observations of Lyman-$\alpha$ (Ly$\alpha$) emitting galaxies \citep{Salvador-Sole2022}. Very massive ($\sim 100-10^{3}\ \rm M_{\odot}$) Pop~III stars are required to produce this double-reionization feature if they evolve normally, while less massive ($\sim 10\ \rm M_\odot$) Pop~III stars can be sufficient if they undergo CHE with boosted ionizing power \citep{Sibony2022}.}. This shows that the fraction of Pop~III stars and their SFE can be constrained by observations of $\tau_0$. However, \citet{Sibony2022} adopt an idealized analytical model for early star formation and reionization that only considers Pop~III stars at $z>15$ and extrapolate the high-$z$ results to predict the full reionization history so that their results should be regarded as upper limits.

In addition to reionization, the UV radiation from Pop~III stars, especially Lyman-band photons, also shapes the 21-cm signal from neutral hydrogen during Cosmic Dawn and the Epoch of Reionization (EoR), which is a promising probe of early structure/star formation at $z\sim 6-30$ \citep[e.g.,][]{Fialkov2013,Fialkov2014rich,Mirocha2019, Schauer2019,Chatterjee2020,Qin2020,Gessey-Jones2022,Kamran2022,Magg2022tr,Munoz2022,Bevins2023,Hassan2023,Mondal2023,Ventura2023,Ventura2025,Fialkov2023,Pochinda2024,Gessey-Jones2025,Dhandha2025}. %Therefore, it is interesting to explore the effects of Pop~III CHE on both reionization and the 21-cm signal in one theoretical framework
In light of this, we extend the previous work by \citet{Sibony2022} to explore the effects of Pop~III CHE on both reionization and the 21-cm signal with state-of-the-art semi-numerical simulations \textsc{21cmSPACE} \citep[e.g.,][]{Visbal2012,Fialkov2012}, which self-consistently model the formation and feedback of Pop~III and Population II (Pop~II) stars and follow (spatially-resolved) evolution of the intergalactic medium (IGM) from $z=50$ down to $z=6$. %the post-reionization epoch ($z=5$).  Gessey-Jones2022,

Beyond stellar UV radiation, X-rays, primarily from X-ray binaries (XRBs), also play important roles in the thermal and ionization evolution of IGM during Cosmic Dawn and EoR \citep[e.g.,][]{Fragos2013,Fialkov2014xray,Pacucci2014,Madau2017,Eide2018,Kaur2022,Gessey-Jones2025}, particularly at $z\lesssim 20$, via heating (and ionization) that can cause the transition of the global 21-cm signal from absorption to emission around $z\sim 10-15$ \citep{Fialkov2014xray}, although theoretical predictions on the contribution of Pop~III XRBs are uncertain especially when CHE is involved. It is shown in \citet{Sartorio2023} that the population-averaged X-ray emission efficiency of XRBs from NR Pop~III stars can be significantly higher (up to a factor of 40) compared with that of Pop~II XRBs \citep{Fragos2013bps,Fragos2013} in optimistic cases \citep[see also][]{Ryu2016}. However, if Pop~III stars undergo CHE, they remain compact throughout their lifetimes, such that very close ($\lesssim 30\ \rm R_{\odot}$) binaries are required to trigger Roche lobe overflow (RLOF), which is the main mechanism of forming Pop~III XRBs identified in \citet{Sartorio2023}. It is still in debate whether such close binaries of Pop~III stars exist as Pop~III binary statistics are highly sensitive to the poorly understood properties of Pop~III star clusters \citep{Liu2021binary}\footnote{In fact, it is found by recent radiative hydrodynamic simulations of primordial star formation \citep[][]{Sugimura2020,Sugimura2023,Park2022,Park2024} that Pop~III protostars tend to migrate outwards due to accretion of gas with high angular momentum, implying that close ($\lesssim 100\ \rm AU$) binaries of massive Pop~III stars are likely rare. If this is true, the X-ray emission from Pop~III RLOF XRBs will be negligible \citep{Liu2021binary}.}. Besides, previous studies focusing on RLOF XRBs find that the donor star must avoid CHE in order to expand and transfer mass to the compact companion \citep{Marchant2017}. %, which is unlikely to happen if most Pop~III stars are born as fast rotators and experience CHE
Here, we make the simplifying assumption that all Pop III stars undergo CHE and cannot produce RLOF XRBs. In reality, hybrid populations of CHE and non-CHE stars are likely more common, and the presence of Pop~III XRBs from RLOF and other mechanisms\footnote{Other than RLOF XRBs, CH Pop~III stars are more likely to produce wind-fed XRBs (that do not require very close binaries) via strong Wolf-Rayet winds \citep{Qiu2019,Zuo2021} or the `Be-phenomenon' in which a fast-rotating star ejects materials to a decretion disk that a compact companion can accrete from \citep{{Reig2011,Rivinius2013,Liu2024}}. However, such scenarios have not been systematically explored for CH stars.} is still possible. Since the focus of this paper is the impact of enhanced UV emission from CHE, we turn off the X-ray emission from Pop~III XRBs for simplicity and adopt an observationally-motivated model for Pop~II XRBs (see Sec.~\ref{sec:setup}). We defer a comprehensive investigation of the UV emission and XRBs from CH Pop~III stars in future work \citep[for the results for NR Pop~III stars, see][]{Gessey-Jones2025}. 
% although Pop~III XRBs may have strong effects on the 21-cm signal \citep[][]{Sartorio2023}. 
% since our focus is the impact of enhanced UV emission from CHE

The paper is organized as follows. In Section~\ref{sec:spec}, we calculate the UV spectra from Pop~III stars with CHE and discuss how they are different from those in the widely-used NR case. In Section~\ref{sec:sim}, we %provide a brief outline of the theory of the 21-cm signal and 
introduce our framework of semi-numerical simulations, with emphasis on the updates in the modelling of ionization with respect to the version used in \citet{Gessey-Jones2023} and the parameter space explored. In Section~\ref{sec:res}, we present our predictions on the 21-cm signal (Sec.~\ref{sec:21cm}) and cosmic ionization history (Sec.~\ref{sec:ion}). Finally, we summarize our findings and discuss their implications in Section~\ref{sec:dis}.

%so that the UV radiation produced will be different. The UV photons from Pop~III stars shape the ionization history and the 21-cm signal from neutral hydrogen at cosmic dawn. 

%what single star evolution can achieve terms of ionising power

 %i.e., Group~III objects according to the classification by \citet[see their fig.~1]{Yoon2016}.}
 
%For instance, the mass distribution, nucleosynthesis and supernova properties of Pop~III stars are recorded in the chemical patterns of second-generation stars formed in the medium enriched by Pop~III stars, which can be observed as extremely metal-poor stars in the local Universe \cite[see e.g.,][]{Yoon2016,Yoon2018,Hansen2019,Dietz2021,Zepeda2022}. The 
%which is also supported by stellar archaeology observations 

%The UV and X-ray photons as well as cosmic rays from the first stars The 21-cm signal from the high-$z$ IGM is shaped by , SNe and galaxies \citep[reviewed by e.g.,][]{Barkana2016}, from which the timing, efficiency and mode of early star formation as well as source properties (e.g., Pop~III IMF) can be constrained .

%

%, to the top-heavy predicted by (magneto)hydrodynamic simulations, , Pop~III stars 

\section{UV Radiation from the first stars}\label{sec:spec}

In this paper, we focus on the UV radiation from the first stars, particularly Lyman-band and ionizing photons, that regulate the 21-cm signal and ionization history during Cosmic Dawn. In this section, we calculate the spectra and ionizing photon production rates of chemically-homogeneous Pop~III stars, which are compared with those for NR Pop~III stars from \citet{Gessey-Jones2022}. The latter include stars (logarithmically-spaced) in the mass range of $M_\star\sim 0.5-500\ \rm M\odot$ evolved by the code \textsc{mesa} \citep[version 12115,][]{Paxton2019} up to the end of MS\footnote{In \citet{Gessey-Jones2022}, the \textsc{mesa} runs of stars with $M_\star\sim 310-500\ \rm M_\odot$ do not reach the end of MS but stop when the star starts to photoevaporate. The subsequent Lyman-band emission is expected to be negligible and ignored (see their appendix a). In this work, we focus on stars below $300\ \rm M_\odot$ whose spectra throughout MS are available.} (i.e., core hydrogen depletion) with no mass loss, whose spectra are derived using the same stellar code adopted here (see below). 
%The UV radiation results are incorporated into semi-numerical simulations of Cosmic Dawn and EoR to predict the effects of Pop~III CHE, as discussed in Sec.~\ref{sec:sim}. 

\begin{figure}
    \centering
    \includegraphics[width=\columnwidth]{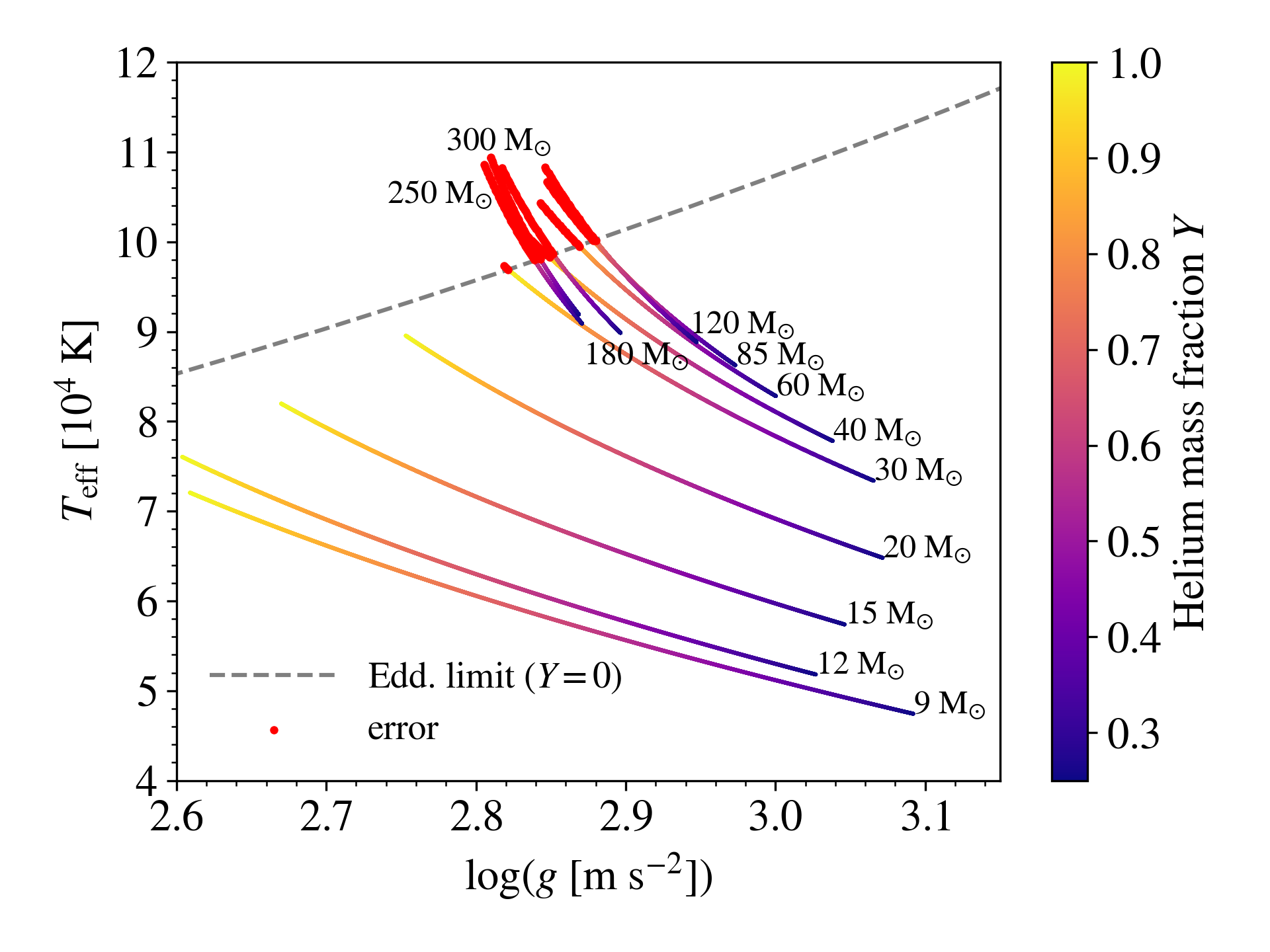}\\
    \vspace{-10pt}
    \includegraphics[width=\columnwidth]{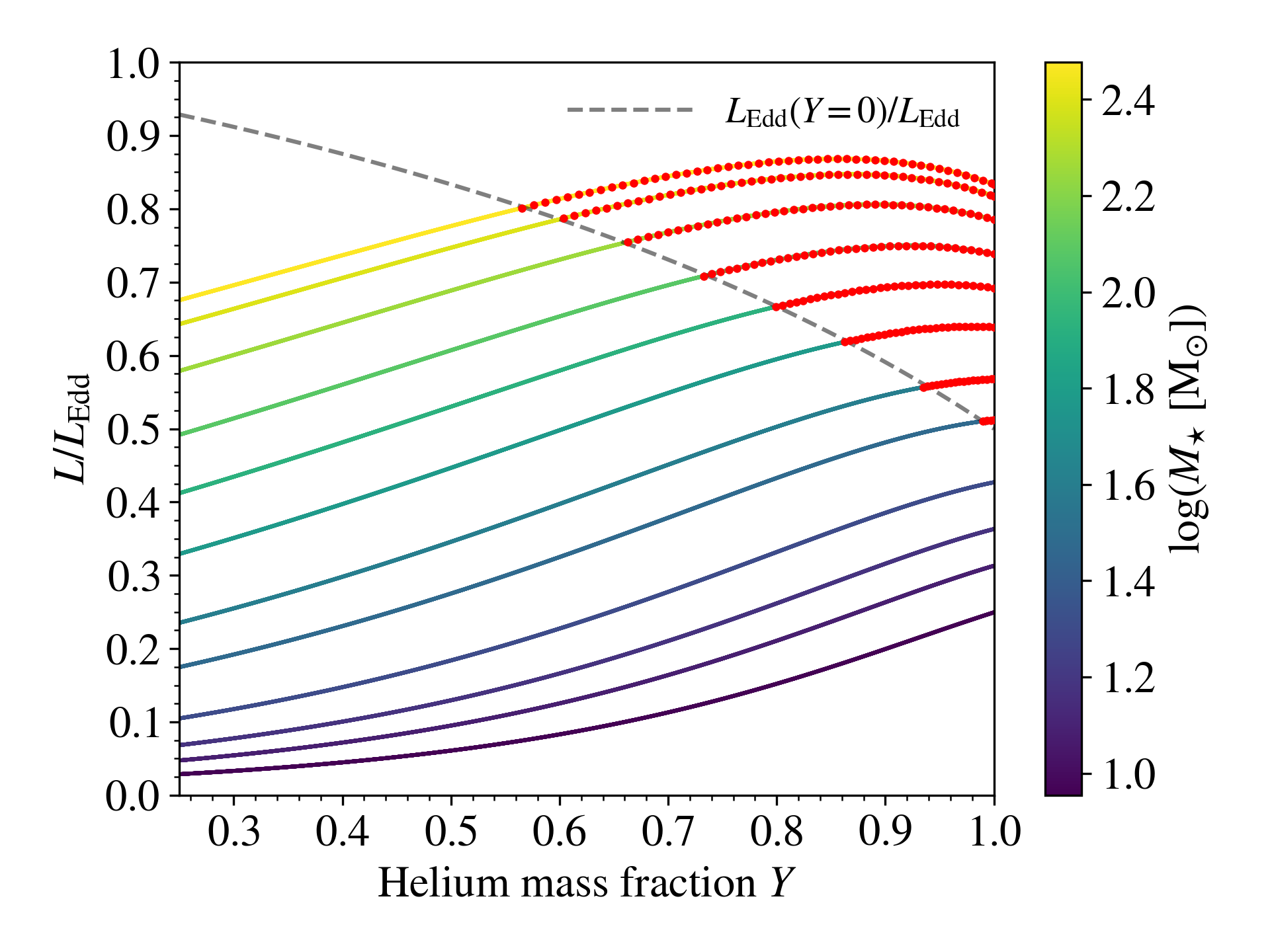}
    \vspace{-20pt}
    \caption{MS evolution of CH Pop~III stars in terms of effective temperature $T_{\rm eff}$ versus logarithm of surface gravity $g$ (\textit{upper}) and Eddington ratio versus helium mass fraction $Y$ (\textit{lower}). In the \textit{upper} panel, the evolution tracks are color coded by $Y$ as an evolution stage indicator. The Eddington limit for atmospheres purely made of hydrogen ($Y=0$) is shown with the dashed curve. In the \textit{lower} panel, the evolution tracks are color coded by stellar mass $M_{\star}$. Now the dashed curve denotes the ratio between the Eddington limit for fully ionized pure hydrogen ($Y=0$) and that for hydrogen-helium plasma with varying $Y$, i.e., $L_{\rm Edd}(Y=0)/L_{\rm Edd}(Y)=\mu_{\rm e}^{-1}=(1-0.75Y)/(1-0.5Y)$. In both panels, the cases where the stellar atmosphere code \textsc{tlusty} fails to converge to a statistic atmosphere are highlighted in red, which occur in late-stage evolution of massive stars ($M_{\star}\gtrsim 30\ \rm M_{\odot}$) when the Eddington limit for pure hydrogen is exceeded. An extrapolation scheme is developed to cover this regime based on the trend of spectra shape evolution in the rest of the parameter space with converged results (see Eq.~\ref{eq:extra} and the text below).}
    \label{fig:teff_logg}
\end{figure}

For simplicity, we only consider (hydrogen-burning) MS that produces the majority of UV photons. 
We derive the MS stellar evolution histories for a grid of 12 (initial) stellar masses $M_{\star}=9$, 12, 15, 20, 30, 40, 60, 85, 120, 180, 250, 300$\ \rm M_\odot$ using the polytrope CHE model detailed in \citet{Sibony2022}, assuming no mass loss. This analytical approach well captures the compact, hot nature of CH stars that is responsible for their enhanced far and extreme UV radiation. The resulting evolution tracks in the HR diagram \citep[fig.~1 in][]{Sibony2022} are consistent with those from detailed stellar evolution simulations \citep[e.g.,][see their fig.~5]{Szecsi2015} for the majority of MS ($Y\lesssim 0.9$). In general, both luminosity and effective temperature increase with time under CHE, while stars without CHE tend to expand significantly and cool down. 
The mass range of Pop~III stars is still in debate in theory, which is expected to vary with the conditions of star formation \citep[e.g.,][]{Liu2024sf}. For simplicity, we adopt a fixed range $M_\star\in [9, 300]\ \rm M_\odot$ that is broadly consistent with observational constraints on the (global average) IMF of Pop~III stars \citep{Hartwig2022,Hartwig2024}. %Note that the impact of CHE on stellar UV emission is stronger for smaller stars as massive stars are hot (radiating at near Eddington rates) anyway \citep{Liu2024che}. %Therefore, if Pop~III IMF extends to lower masses. 

As an improvement of the previous results in \citet{Sibony2022} which adopted black-body spectra, we use the stellar atmosphere code \sacode\footnote{We use \sacode because it has been successfully applied to massive (Pop~III) stars \citep{Schaerer2002,Lanz2003,Lanz2007,Gessey-Jones2022}. In particular, the spectra of NR Pop~III stars from \citet{Gessey-Jones2022}, which serve as the reference to compare our CHE results with, are computed using \sacode. }\citep[version 205,][]{Hubeny1988,Hubeny2017a,Hubeny2017b,Hubeny2017c,Hubeny2021} to calculate the spectra of each star at several time steps\footnote{The time steps are chosen such that the variations in $\log(T_{\rm eff})$ and $\log(g)$ are less than 0.01 dex at each step.} throughout MS given the corresponding effective temperature $T_{\rm eff}$, surface gravity $g$, and chemical abundances. Here we ignore the trace amount of metals synthesized during MS, so that the chemical composition of the atmosphere is completely determined by the (overall) helium mass fraction $Y$ of a CH Pop~III star. 
The top panel of Fig.~\ref{fig:teff_logg} shows the evolution tracks of the 12 stars in the space of $T_{\rm eff}$ and $\log(g)$, colour-coded by $Y$, which serves as a
good evolution stage indicator, since our CH stars burn all hydrogen into helium, reaching $Y=1$ at the end of MS. Contrary to the case of NR Pop~III stars (see fig.~2 in \citealt{Gessey-Jones2022} and fig.~1 in \citealt{Sibony2022}) where $T_{\rm eff}$ generally decreases with time during MS, CH Pop~III stars become hotter as $Y$ increases for $M_{\star}\sim 9-300\ \rm M_\odot$. The reason is that CH stars remain compact (i.e., very little expansion)\footnote{For $M_{\star}\sim 9-300\ \rm M_\odot$, the stellar radii of CH Pop~III stars increase by $\Delta R_{\star}= 0.9\pm 0.2\ \rm R_\odot$ during MS, almost independent of $M_{\star}$, while those of NR Pop~III stars increase by $\Delta R_{\star}\sim 2-100\ \rm R_\odot$, reaching up to $\sim 10$ times of the zero-age MS value (see table~1 in \citealt{Sibony2022}).} despite the luminosity enhancement from increasing $Y$.
%The evolution of the 12 CH stars are shown in the top panel of Fig.~\ref{fig:teff_logg}.\citealt{Tanikawa2020}

\begin{figure}
    \centering
    \includegraphics[width=\columnwidth]{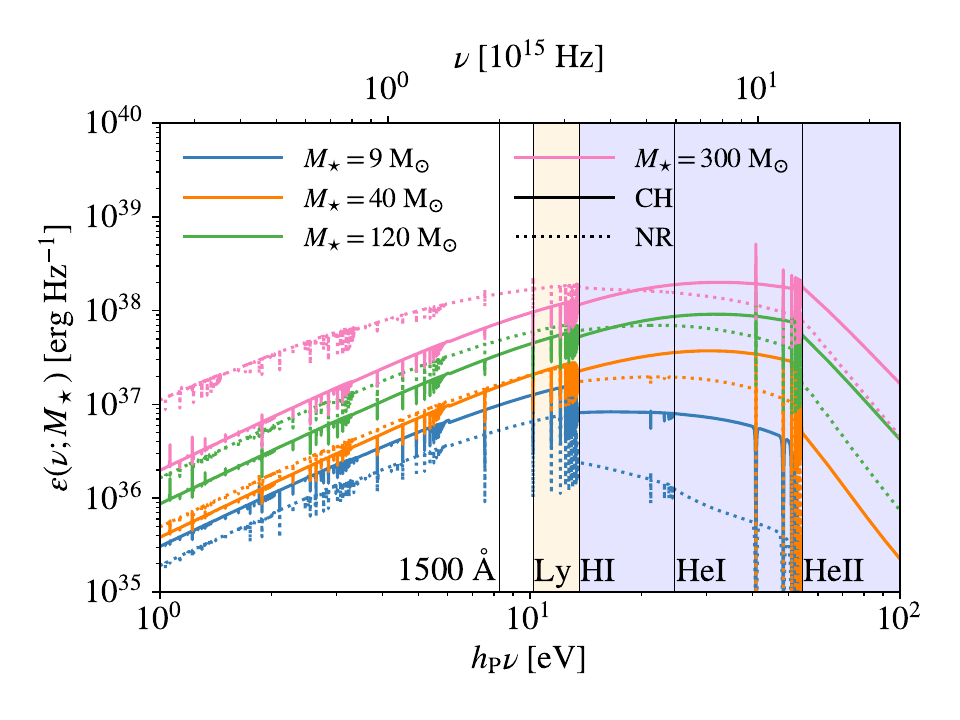}\\
    \vspace{-10pt}
    \includegraphics[width=\columnwidth]{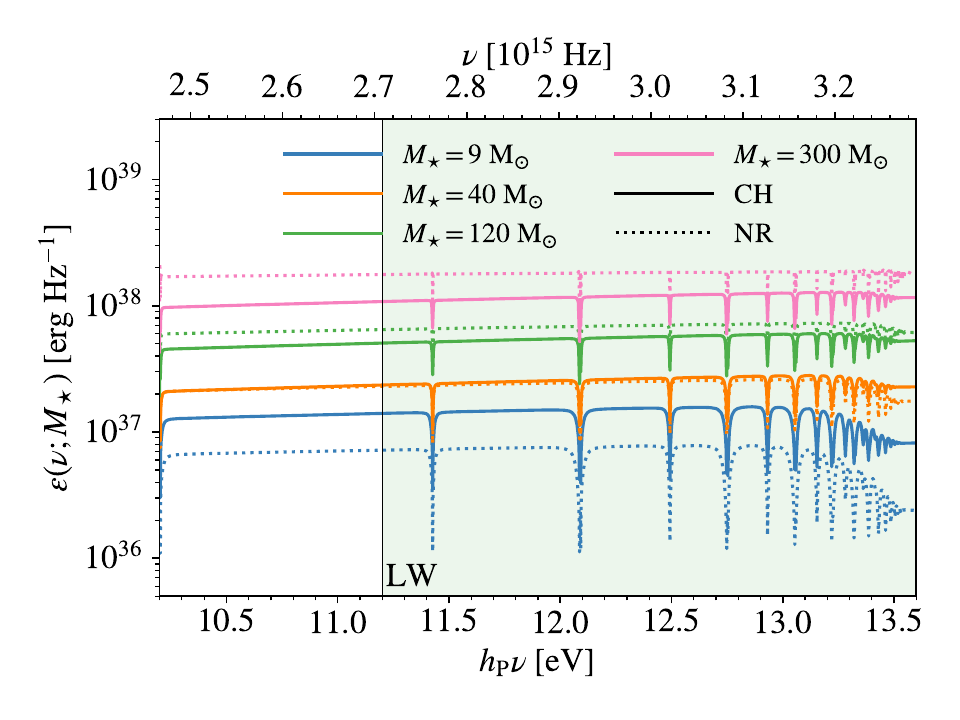}
    \vspace{-20pt}
    \caption{MS lifetime-integrated spectra of CH (solid) and NR (dotted) Pop~III stars with masses of $M_{\star}=9$, 40, 120 and 300~$\rm M_\odot$ (from bottom to top). The \textit{upper} panel show the spectra in the photon energy range of $h_{\rm P}\nu\in [1,100]$~eV, where the vertical lines label the characteristic energies for Lyman-band (Ly, $h_{\rm P}\nu\sim 10.2-13.6$~eV, orange shaded area), hydrogen (HI, $h_{\rm P}\nu>13.6$~eV), helium first (HeI, $h_{\rm P}\nu>24.6$~eV) and second (HeII, $h_{\rm P}\nu>54.4$~eV) ionizing photons, and rest-frame UV at $1500$\ \AA\ corresponding to $h_{\rm P}\nu\simeq 8.3$~eV. The spectrum at $h_{\rm P}\nu>13.6$~eV (blue shaded area) is integrated to calculate the production rate of (hydrogen) ionizing photons. The \textit{lower} panel zooms into the Lyman band ($h_{\rm P}\nu\sim 10.2-13.6$~eV), where the energy/frequency range of $\rm H_2$-dissociating Lyman-Werner (LW) photons ($h_{\rm P}\nu\sim 11.2-13.6$~eV) are highlighted by the shaded area. The spectra of NR stars are derived by interpolation (over mass) of the spectra in \citet{Gessey-Jones2022}. We use $h_{\rm P}$ to denote the Planck constant to avoid confusion with the Hubble constant parameter $h$ in cosmology.}
    \label{fig:spec}
\end{figure}
% cause numerical problems

At each point on the evolution track chosen for spectrum calculation, we adopt the \sacode settings and iterative procedure decribed in \citet[see their Sec.~3.4]{Gessey-Jones2022} to search for a converged atmosphere model\footnote{Convergence is achieved when the variations of all atmosphere properties are less than 1\% for an iteration in \sacode.}. Converged results are obtained for most cases in the relevant parameter space with $g\sim 10^{2.6-3.1}\ \rm m\ s^{-2}$ and $T_{\rm eff}\lesssim 9.6\times 10^{4}$~K, while \sacode fails for the hottest atmospheres that occur in late-stage ($Y\gtrsim 0.6$) evolution of massive ($M_{\star}\gtrsim 30\ \rm M_\odot$) CH stars. After close scrutiny, we find that these atmospheres will be super Eddington if they are purely made of hydrogen, i.e., the radiation pressure from Thomson scattering for ionized hydrogen exceeds the gravity at the stellar surface. However, considering the high abundances of helium mixed into them during CHE, these atmospheres will approach but still remain below the Eddington limit\footnote{The Eddington luminosity $L_{\rm Edd}$ of fully ionized hydrogen-helium plasma is proportional to the number of electrons per baryon $\mu_{\rm e}=(1-0.5Y)/(1-0.75Y)$ which increases with the helium mass fraction $Y$.}, as shown in the lower panel of Fig.~\ref{fig:teff_logg}. It seems that \sacode always assumes the existence of a layer of pure hydrogen in the atmosphere, and the computation fails to converge to a static atmosphere when this hydrogen layer becomes super Eddington. 
In reality, the atmospheres of massive %($M_{\star}\gtrsim 100\ \rm M_{\odot}$) 
fast-rotating CH Pop~III stars may indeed be unstable when they approach the Eddington limit by the end of MS ($Y\gtrsim 0.6$), such that %render the atmosphere vulnerable. 
they can enter the Wolf-Rayet phase with non-negligible mass loss (a few percent of the initial mass) enhanced by fast rotation and surface enrichment of heavy elements (e.g., C, N, O, and Fe) that boosts the opacity of the atmosphere \citep[e.g.,][]{Jeena2023}. 
This scenario is not considered in our analytical CHE model. %, and we defer a more detailed investigation of this regime to future work. 
Here, we assume that mass loss is negligible for CH Pop~III stars throughout MS, i.e., until complete hydrogen depletion ($Y=1$). %{\color{red}Such mass loss can potentially reduce the UV luminosity of a massive star in the late stage of evolution \citep[e.g.,][]{Kubatova2019}. Therefore, our results should be regarded as optimistic estimates for the impact of CHE}. 
Under this assumption, we develop a simple extrapolation scheme to compute the late-stage spectra of the 8 stars with $M_{\star}\ge 30\ \rm M_\odot$ for which \sacode results are unavailable, as discussed in Appendix~\ref{apdx:ext}. In reality, mass loss can reduce the UV luminosity of a massive star, especially for the late evolution stage. However, it is shown in \citet[see their fig.~2]{Kubatova2019} that even given the maximum effective surface metallicity $Z\sim 10^{-4}$ achievable in Pop~III CH stars \citep{Jeena2023} from the beginning, the impact of Wolf-Rayet-like winds on the spectra of CH stars with $M_\star\sim 20-130\ \rm M_\odot$ at $h_{\rm p}\nu\sim 8-54.4$~eV is always negligible for the majority of MS ($Y\lesssim 0.98$) under various wind prescriptions. The ignorance of mass loss does not change our conclusions thanks to the metal-free/poor nature of Pop~III stars. In contrast, for metal-rich stars with $Z\gtrsim 0.002$, CHE can trigger strong mass loss via optically thick winds that significantly affect the evolution during MS (peeling off the hydrogen-rich envelope), which is required to reproduce the single Wolf-Rayet stars observed in the Small Magellanic Cloud \citep{Boco2025}.

Since we only consider massive ($M_{\star}\ge 9\ \rm M_\odot$) stars with short MS lifetimes $t_{\rm MS}\lesssim 30\ \rm Myr$, their emission can be modelled as instantaneous. It is shown in \citet[see their table~1 and fig.~B1]{Gessey-Jones2022} for NR Pop~III stars that adopting the instantaneous emission approximation has negligible impact on the 21-cm signal when the Pop~III IMF is dominated by short-lived ($\lesssim 30\ \rm Myr$) massive stars. We have verified by numerical experiments that this is also the case for our CH Pop~III stars. Therefore, now we focus on the MS lifetime-integrated spectrum $\epsilon(\nu;M_\star)=\int_{0}^{t_{\rm MS}} L_{\nu}(t)dt$ to demonstrate the difference between CH and NR Pop~III stars, given the specific luminosity $L_{\nu}(t)$. 
Fig.~\ref{fig:spec} shows the spectra for CH Pop~III stars with $M_{\star}=9$, 40, 120 and 300$\ \rm M_\odot$ in comparison with those for NR Pop~III stars based on linear interpolation of the spectra from \citet{Gessey-Jones2022} in the $\log\epsilon$-$\log M_{\star}$ space. We find that CH stars have harder spectra due to their compact, hot nature, with boosted emission of ionizing photons, especially for relatively small stars. For both CH and NR Pop~III stars, the Lyman-band continuum is almost flat (see the lower panel in Fig.~\ref{fig:spec}), which is enhanced by CHE for $M_{\star}\lesssim 50\ \rm M_\odot$ but reduced for $M_{\star}\gtrsim 50\ \rm M_\odot$. The trend at $M_{\star}\gtrsim 50\ \rm M_\odot$ is caused by the fact that massive stars under CHE are so hot that they primarily produce more energetic (ionizing) photons and less Lyman-band photons. 

% \footnote{$\epsilon(\nu;M_\star)$ is related to the photon-number emissivity per stellar baryon $\epsilon_{\rm b}(\nu)$ in the notation of \citet[see their fig.~7]{Gessey-Jones2022} through $\epsilon_{\rm b}(\nu)=\epsilon(\nu;M_\star)(\mu m_{\rm H}/M_\star)/(h_{\rm P}\nu)$, where $\mu\simeq 1.22$ is the mean molecular weight of primordial gas and $m_{\rm H}$ is proton mass.}

\begin{comment}
\begin{figure}
    \centering
    \includegraphics[width=\columnwidth]{lifetime_spectra.pdf}
    \vspace{-20pt}
    \caption{Same as Fig.~\ref{fig:spec_all} but zoomed into the Lyman band.}
    \label{fig:spec_ly}
\end{figure}
\end{comment}

\begin{figure}
    \centering
    \includegraphics[width=\columnwidth]{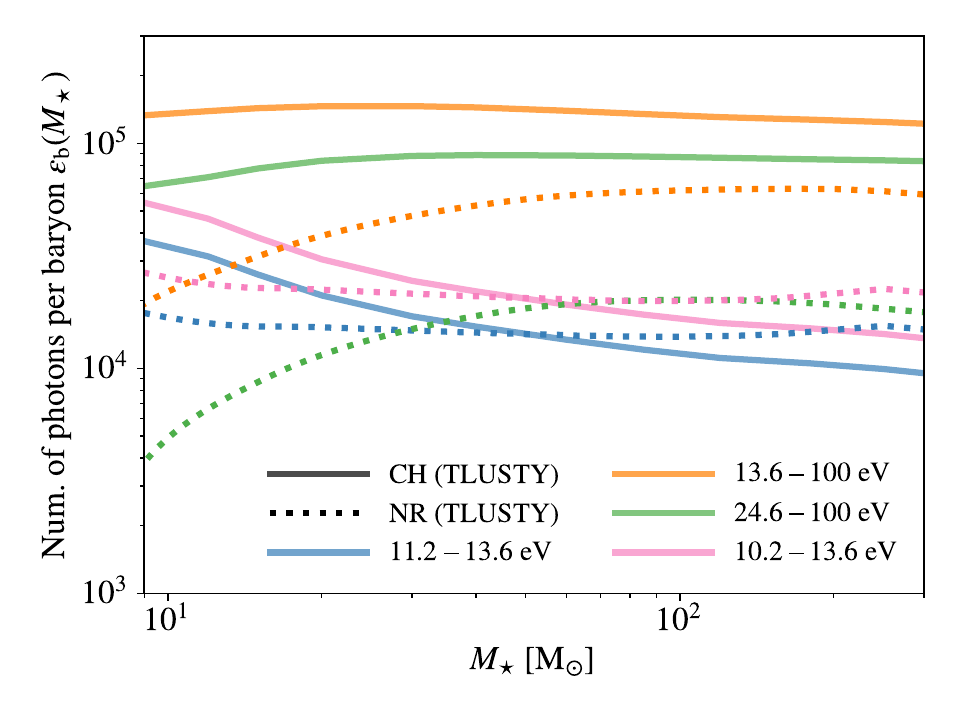}
    \vspace{-20pt}
    \caption{Number of photons emitted per stellar baryon $\epsilon_{\rm b}(M_\star)$ during MS as a function of stellar mass $M_{\star}$ based on \sacode stellar atmosphere models in 4 bands: LW ($11.2-13.6$~eV, blue), ionizing radiation for hydrogen ($13.6-100$~eV, orange) and helium ($24.6-100$~eV, green), and Lyman-band ($10.2-13.6$~eV, pink). The results for CH and NR stars are shown with solid and dotted curves, respectively. The latter are derived from the spectra in \citet{Gessey-Jones2022}.}
    \label{fig:nphoton}
\end{figure}

% $m_{\rm H}$ is proton mass

{To better quantify the difference made by CHE, we calculate the number of photons emitted per stellar baryon (i.e., emission efficiency) during MS as a function of $M_{\star}$: 
\begin{align}
    \epsilon_{\rm b}(M_\star)=(m_{\rm H}/M_\star)\int_{\nu_1}^{\nu_2}\epsilon(\nu;M_{\star})d\nu\ ,\label{epsilon_b}
\end{align}
where $\nu_1$ and $\nu_2$ define the frequency range of the band of interest, and $m_{\rm H}$ is proton mass.}
%$\epsilon_{\rm b}(M_\star)=(m_{\rm H}/M_\star)\int\epsilon(\nu;M_{\star})d\nu$ 
We consider 4 representative bands relevant for reionization and the 21-cm signal (see Sec.~\ref{sec:sim}): Lyman-band ($10.2-13.6$~eV), Lyman-Werner (LW, $11.2-13.6$~eV), hydrogen ($13.6-100$~eV) and helium first ($24.6-100$~eV) ionizing photons. Our analysis is restricted to UV photons below 100~eV because the number of more energetic photons is negligible. In Fig.~\ref{fig:nphoton} we compare the results for CH Pop~III stars and those for NR Pop~III stars \citep{Gessey-Jones2022}. According to \sacode results + extrapolation, with CHE, the emission efficiency of LW photons $\epsilon_{\rm b}^{\rm Ly}(M_\star)$ decreases with increasing $M_{\star}$, while $\epsilon_{\rm b}^{\rm Ly}(M_\star)\sim 1.5\times 10^{4}$ is almost constant for $M_{\star}\sim 9-300\ \rm M_\odot$ in the NR case. It turns out that $\epsilon_{\rm b}^{\rm Ly}(M_\star)$ is enhanced by CHE for $M_{\star}\lesssim 50\ \rm M_\odot$ by up to a factor of 2 at $M_{\star}=9\ \rm M_\odot$ but reduced for $M_{\star}\gtrsim 50\ \rm M_\odot$ by up to $36\%$ at $M_{\star}=300\ \rm M_\odot$. The results for the full Lyman band are similar to those for LW photons. On the other hand, the production of ionizing photons is significantly enhanced with CHE across the mass range considered here, by a factor of $\sim2-6$ and $\sim 3-14$ for hydrogen and helium first ionization, respectively. The boost is stronger for more energetic photons and less massive stars. We compare our results based on \sacode with those from black-body spectra in Appendix~\ref{apdx:bb}, which shows that the black-body approximation causes non-negligible errors in $\epsilon_{\rm b}(M_\star)$ by up to a factor of a few, although the general trends still hold. %We only use \sacode results in the semi-numerical simulations introduced below, for both CH and NR Pop~III stars. 

%Here we focus on the mass range $M_{\star}$

%restrict our analysis to the (initial) stellar mass range $M_{\star}\in []$

%We combine the \textsc{tlusty} stellar atmosphere model (version 205) with 

%Stellar evolution models: non-rotating (NR) Pop~III stars \citep{Gessey-Jones2022}, chemically homogeneous (CH) Pop~III stars \citep{Sibony2022}. Stellar atmosphere model: \textsc{tlusty} version 205 \citep{Hubeny1988,Hubeny2017a,Hubeny2017b,Hubeny2017c,Hubeny2021}.

%$ in \citet{Gessey-Jones2022,Gessey-Jones2023}\citet[see their sec.~3.1, 3.2, and references therein]{Gessey-Jones2023}

\section{Semi-numerical simulations of Cosmic Dawn and EoR}\label{sec:sim}
We use the semi-numerical code \textsc{21cmSPACE} \citep[e.g.,][]{Visbal2012,Fialkov2012} to simulate the 21-cm signal and ionization history of the IGM at $z\ge 6$. In this section, we briefly summarize the key physical elements in our numerical framework %focusing on the roles played by Pop~III stars. 
particularly focusing on the prescription for IGM ionization by stellar UV radiation, which is updated to self-consistently capture the effects of CH/NR Pop~III stars and photo-heating feedback by IGM ionization (Sec.~\ref{sec:ion_model}). The reader is referred to \citet[see their sec.~4]{Gessey-Jones2022} and \citet[see their sec.~3.1, 3.2, and references therein]{Gessey-Jones2023} for more detailed descriptions of the implementations of the relevant physics.

\subsection{Theoretical background of the 21-cm signal}

The 21-cm signal is produced by the absorption or emission at the 21-cm spectral line ($\nu_{21}=1420\ \rm MHz$) of hydrogen atoms via transitions between hyper-fine states. It is measured by the differential 21-cm brightness temperature seen by an observer at $z=0$:
\begin{align}
    T_{21}(\nu)=(1-e^{-\tau_{21}})\frac{T_{\rm S}(z)-T_{\gamma}(z)}{(1+z)}\ ,
\end{align}
which denotes the difference in radiation temperature at the observer-frame frequency $\nu=\nu_{21}/(1+z)$ caused by IGM absorption/emission at redshift $z$, given the radio background temperature $T_{\gamma}(z)$ at $\nu_{21}$. Here $T_{\rm S}(z)$ is the spin temperature that reflects the occupation fractions of hydrogen hyper-fine states, and $\tau_{21}$ is the 21-cm optical depth \citep{Madau1997,Barkana2016}: 
\begin{align}
    \tau_{21}(z)=\frac{3hc^{3}A_{10}}{32\pi k_{\rm B}\nu_{21}^{2}}\frac{x_{\rm HI}(z)n_{\rm H}(z)}{(1+z)(dv_{\lVert}/dr_{\lVert})}\frac{1}{T_{\rm S}(z)}\ ,\label{tau_21}
\end{align}
where $A_{10}=2.85\times 10^{-15}\ \rm s^{-1}$ is the spontaneous emission rate of the 21-cm transition, $dv_{\lVert}/d r_{\lVert}\sim H(z)/(1+z)$ is the proper velocity gradient along the line of sight caused by cosmic expansion given the Hubble parameter $H(z)$, $n_{\rm H}(z)$ is the physical number density of hydrogen nuclei, and $x_{\rm HI}(z)$ is the neutral fraction. For simplicity, we assume that $T_{\gamma}$ is the radiation temperature of the cosmic microwave background (CMB) as $T_{\gamma}(z)=2.725(1+z)\ \rm K$ (see, e.g., \citealt{Ewall-Wice2018,Feng2018,Fialkov2019,Reis2020} for models of excess radio backgrounds). 

Clearly, the 21-cm signal is sensitive to the redshift evolution of $T_{\rm S}$, $x_{\rm HI}$, and $n_{\rm HI}$ in the IGM, which are regulated by cosmic structure formation, star formation, and radiation transfer. In particular, the spin temperature $T_{\rm S}$ is determined by three processes: CMB scattering, atomic collisions, and Ly$\alpha$ scattering, i.e., the Wouthuysen–Field (WF) effect \citep{Wouthuysen1952,Field1958}. The strengths of these processes are denoted by the corresponding coupling coefficients, $x_\gamma$, $x_{\rm c}$, and $x_{\alpha}$, derived from atomic physics. In the end, $T_{\rm S}$ can be written as \citep[e.g.,][]{Field1958,Madau1997,Barkana2016}\footnote{In practice, we use $x_\gamma=1$ for simplicity, corresponding to optimal coupling in the optically thin limit ($\tau_{21}\rightarrow0$).}
\begin{align}
    T_{\rm S}^{-1}=\frac{x_\gamma T_{\gamma}^{-1}+x_{\rm c}T_{\rm K}^{-1}+x_{\alpha}T_{\rm C}^{-1}}{x_\gamma+x_{\rm c}+x_{\alpha}}\ .\label{Ts}
\end{align}
Here, $T_{\rm K}$ is the kinetic temperature of the IGM, and $T_{\rm C}$ is the colour temperature of Ly$\alpha$ radiation \citep{Madau1997,Barkana2016}. In the neutral IGM before reionization that is highly opaque to resonant scattering of Ly$\alpha$ photons, $T_{\rm C}$ is very close to $T_{\rm K}$ \citep{Field1959}. %These coefficients are calculated from atomic physics and the intensity of Lyman-band radiation. 
A variety of processes are considered to drive the evolution of $T_{\rm K}$ \citep[see eq.~13 in][]{Gessey-Jones2023}, including cosmic expansion, structure formation, ionization, X-ray heating, Compton heating, and Ly$\alpha$ heating \citep[e.g.,][]{Madau1997,Fialkov2014xray,Reis2021}.
%cosmic expansion, structure formation, ionization, X-ray heating \citep{Fialkov2014xray}, Compton heating \citep{Madau1997}, and Ly$\alpha$ heating \citep{Reis2021}. %, and radio background heating \citep{Venumadhav2018,Fialkov2019,Reis2021}. 
%(Venumadhav et al. 2018; Fialkov \& Barkana 2019; Reis et al. 2021).

Shortly after the onset of first star formation, the UV radiation from Pop~III stars is expected to make the WF effect dominant over the other processes, leading to $T_{\rm S}\approx T_{\rm K}$. The relevant coupling coefficient $x_{\alpha}$ is proportional to the intensity $J_{\alpha}$ of Ly$\alpha$ photons. Following the methodology of \citet{Reis2021} as an extension of \citet{Barkana2005} and \citet{Fialkov2014}, $J_{\alpha}$ is calculated by convolving radiation transfer window functions with the Lyman-band emissivity fields in our simulations, which are then derived from the spectra and star formation histories of individual stellar populations \citep[][see their eq.~14 and the following text]{Gessey-Jones2023}. For the Pop~III emissivity field, we follow the method in \citet{Gessey-Jones2022}, integrating the spectra of individual stars (Fig~\ref{fig:spec}) over the IMF, which is assumed to follow a power-law form $dN/dM_\star\propto M_\star^{-\alpha}$ in a fixed range $M_\star\in [9,300]\ \rm M_\odot$ for both NR and CH stars. We consider 3 values of the slope $\alpha=0$, 1, and 2.35 to explore the effects of IMF. %, which are expected to be small \citep{Gessey-Jones2022}. 
The star formation histories of Pop~III and II stars are also used to model IGM ionization, as discussed below.

\subsection{Updated prescription for IGM ionization}\label{sec:ion_model}

We derive the IGM neutral fraction $x_{\rm HI}$ of each cell in the simulation box under the assumption that (1) the ionization from stellar UV radiation takes the form of fully ionized bubbles, while (2) other sources of ionization (e.g., X-rays) can travel long distances in the IGM to cause partial ionization. For the former process, the excursion set formalism \citep{Furlanetto2004,Mesinger2011} is used to identify fully ionized regions: a cell at position $\bm{x}$ is fully ionized, i.e., $x_{\rm HI}(\bm{x})=0$, if there exists a spherical volume of radius $R$ centred on it in which the time-integrated (effective) number of ionizing UV photons exceeds the number of neutral atoms, i.e.
\begin{align}
    \exists R<R_{\max}\ ,\quad \text{so that}\quad n_{\gamma,\rm ion}(\bm{x},R)>1-x_{\rm e,oth}(\bm{x},R)\ ,\label{ion}
\end{align}
where $R_{\max}$ is the maximum radius of ionized bubbles, $n_{\gamma,\rm ion}(\bm{x},R)$ is the cumulative effective number of ionizing photons per baryon averaged over a sphere of radius $R$ centred on $\bm{x}$, and $x_{\rm e,oth}(\bm{x},R)$ is the partially ionized fraction caused by long-range agents (2) averaged over the same sphere, which is evolved using eq.~17 in \citet{Gessey-Jones2023}. We adopt $R_{\max}=50\ \rm cMpc$ motivated by theoretical predictions and observations of the mean free path of ionizing photons at the end of reionization \citep[$z\sim 5-6$,][]{Wyithe2004,Furlanetto2005,Lewis2022,Zhu2023}. 

If the criterion in Eq.~\ref{ion} is never satisfied down to the resolution (i.e., cell size $\Delta x$) of the simulation, the ionized bubble is confined in the cell. In this case, the cell is treated as a two-phase medium, one fully ionized with a volume occupation fraction of $n_{\gamma,\rm ion}(\bm{x})$ and one ionized to $x_{\rm e,oth}(\bm{x})$, which is an approximation validated by \citet{Zahn2011}. The overall neutral fraction is then calculated locally as 
\begin{align}
    x_{\rm HI}(\bm{x})=1-n_{\gamma,\rm ion}(\bm{x})-[1-n_{\gamma,\rm ion}(\bm{x})]x_{\rm e,oth}(\bm{x})\ .
\end{align}
Here, following previous studies \citep{Fialkov2014rich,Fialkov2017,Gessey-Jones2023}, the single ionization fractions of hydrogen and helium are assumed to be identical, and helium double ionization in the IGM is not modelled in the simulation, which is expected to be driven by active galactic nuclei at relatively low redshifts $z\sim 3-6$ \citep[see their fig.~8]{Gotberg2020} beyond the scope of this paper\footnote{In primordial nebulae hosting massive Pop~III stars, helium double ionization can be efficient locally due to the hard stellar spectra, which powers strong HeII$\lambda$1640\AA\ lines as a key signature of Pop~III star formation \citep[e.g.,][]{Bromm2001,Venditti2024,Lecroq2025}. However, the number of helium doubly ionizing photons from Pop~III stars is too low to have a strong effect on cosmic reionization even under CHE \citep{Sibony2022}. }. Besides, we assume that once a region is fully ionized, it remains ionized thereafter. That is to say, ionized bubbles are not allowed to shrink by recombination in our simulations although this may happen in reality if the UV emissivity declines, which can even cause non-monotonic evolution of the cosmic average neutral fraction \citep{Salvador-Sole2017,Salvador-Sole2022}. We defer a more comprehensive model for the dynamical balance between ionization and recombination to future work.

In this work, we introduce a new scheme to calculate $n_{\gamma,\rm ion}(\bm{x},R)$ considering the Pop~III and II contributions separately. This is an improvement over \citet{Gessey-Jones2023} in which $n_{\gamma,\rm ion}(\bm{x},R)$ is associated to the fraction of baryons collapsed into galaxies using a phenomenological efficiency parameter $\zeta$ without distinguishing Pop~III and II stars. To be specific, we have
\begin{align}
\begin{split}
    &n_{\gamma,\rm ion}(\bm{x},R)=\min[1,n_{\gamma,\rm ion,II}(\bm{x},R)+n_{\gamma,\rm ion,III}(\bm{x},R)]\ ,\\ &n_{\gamma,{\rm ion},i}(\bm{x},R)\equiv \frac{\mu\epsilon_{{\rm b},i}^{\rm ion}f_{{\rm esc},i}f_{{\rm stellar},i}(\bm{x},R)}{(1+N_{\rm rec})}\ , \quad i=\rm II,\ III\ ,
\end{split}
\label{nion}
\end{align} %+\frac{\epsilon_{\rm b,III}^{\rm ion}f_{\rm esc,III}f_{\rm stellar,III}}{(1+N_{\rm rec})}
where $N_{\rm rec}$ is the average number of recombinations experienced per baryon before remaining ionized %to be maintained fully ionized 
(which captures the effects of IGM clumping and attenuation), %\footnote{\color{red}The effective clumping factor as the ratio of the volume-averaged recombination rate and the recombination rate at the cosmic average density can be written as $C_{\rm eff}=1+N_{\rm rec}$.}, 
$f_{{\rm stellar},i}(\bm{x},R)$ is the fraction of \textit{all} baryons that have formed Pop~$i$ stars\footnote{Note that $f_{\rm stellar}$ and $f_{\star}$ are distinct. The latter generally describes the fraction of gas \textit{in star-forming halos} that become stars (SFE), while $f_{\rm stellar}$ is defined for \textit{all} baryons including those not in star-forming haloes.}, and $\epsilon_{{\rm b},i}^{\rm ion}$ and $f_{{\rm esc},i}$ are the corresponding (IMF-averaged) number of ionizing photons produced per stellar baryon and escape fraction. We calculate $f_{{\rm stellar},i}$ by integrating the star formation rate density (SFRD) $\dot{\rho}_{\star,i}$:
\begin{align}
    f_{{\rm stellar},i}(\bm{x},R,z)=\frac{\int_{z}^{z_{\rm ini}}\dot{\rho}_{\star,i}(\bm{x},R,z')|dt/dz'|dz'}{\rho_{\rm b}(\bm{x},R,z)}\ ,\ i=\rm II,\ III\ ,\label{fgas}
\end{align}
where $z_{\rm ini}=50$ is the initial redshift of the simulation and $\rho_{\rm b}(\bm{x},R,z)$ is the baryon density. Note that all the quantities involved in Eqs.~\ref{nion} and \ref{fgas} are averaged over spherical volumes in the excursion-set formalism (Eq.~\ref{ion}). 

The SFRD in each cell plays a crucial role in the above calculation. For cells that are not fully ionized by UV radiation, we treat the contributions from fully ionized and partially ionized regions differently, which are then combined with the weights $n_{\gamma,\rm ion}$ (ionized fraction of the cell) and $1-n_{\gamma,\rm ion}$. For Pop~III stars, we estimate the SFRD under the assumption that Pop~III stars only form in one burst (per halo) once the halo crosses a mass threshold $M_{\min}$, which is set to the minimum of the critical masses for efficient molecular ($M_{\rm mol}$) and atomic cooling ($M_{\rm atm}$), i.e., $M_{\min}=M_{\rm cool}\equiv\min(M_{\rm mol},M_{\rm atm})$ \citep{Magg2022tr}. Here, $M_{\rm mol}= M_{z=20}f_{\rm LW}f_{v_{\rm bc}}[(1+z)/21]^{-3/2}$ with suppression factors for LW feedback ($f_{\rm LW}$) and streaming motion ($f_{v_{\rm bc}}$) given by the simulation-motivated fitting formula detailed in \citet[see their sec. 2.2.2]{Munoz2022}. %\citet[see their sec. 3.2]{Gessey-Jones2023}. 
We adopt $M_{z=20}=5.8\times 10^{5}\ \rm M_\odot$ \citep{Schauer2021} and $M_{\rm atm}=10^8\ {\rm M_\odot}[V_{\rm c,atm}/(17\ {\rm km\ s^{-1}})]^{3}[(1+z)/10]^{-3/2}$ given $V_{\rm c,atm}=16.5\ \rm km\ s^{-1}$, which is the circular velocity of critical atomic-cooling haloes \citep{Gessey-Jones2022}. For simplicity, Pop~III star formation is forbidden in fully ionized regions (i.e., fully ionized cells and UV-ionized bubbles in partially ionized cells), since haloes massive enough ($\gtrsim 10^9\ \rm M_\odot$) to overcome photo-heating feedback (see below) are typically metal-enriched and no longer form Pop~III stars \citep[see, e.g., figs.~6 and 9 in][]{Ventura2023}\footnote{Some cosmological simulations found that Pop~III star formation can still occur in pockets of pristine gas in massive haloes under inhomogeneous metal enrichment \citep[e.g.,][]{Tornatore2007,Xu2016,Benitez-Llambay2020,Liu2020,Venditti2023}. How much this scenario contributes to the overall Pop~III SFRD is still uncertain as metal mixing is poorly understood.}. In this way, the Pop~III SFRD in a cell at cosmic age $t$ is given by
\begin{comment}
\begin{align}
\begin{split}
    &\dot{\rho}_{\star,\rm III}(t)=\frac{f_{\star,\rm III}f_{\rm b}}{\Delta t_{\star,\rm III}}\max\left[0,M_{\min}(t)\int_{M_{\min}(t)}^{M_{\max}}\frac{dn_{\rm h}(t)}{dM'}dM'\right.\\
    &\left.-M_{\min}(t-\Delta t_{\star,\rm III})\int_{M_{\min}(t-\Delta t_{\star,\rm III})}^{M_{\max}}\frac{dn_{\rm h}(t-\Delta t_{\star,\rm III})}{dM'}dM'\right]\ .    
\end{split}
\label{sfr_3}
\end{align}
\end{comment}
\begin{align}
\begin{split}
    &\dot{\rho}_{\star,\rm III}(t)=\frac{f_{\star,\rm III}(1-n_{\gamma,\rm ion})}{\Delta t_{\star,\rm III}}\max\biggl\{0,\frac{ }{ }M_{\min}(t)\times \\
    &\int_{M_{\min}(t)}^{M_{\max}}\left[\frac{d\rho_{\rm g}(t+0.5\Delta t_{\star,\rm III})}{dM'}-\frac{d\rho_{\rm g}(t-0.5\Delta t_{\star,\rm III})}{dM'}\right]\frac{dM'}{M'}\biggr\} .
\end{split}
\label{sfr_3}
\end{align}
Here $d\rho_{\rm g}/dM'\equiv (1+\delta)f_{\rm b}M'dn_{\rm h}/dM'$ is the baryon density per unit halo mass, given %$f_{\rm b}=\Omega_{\rm b}/\Omega_{\rm m}\simeq 0.16$ \citep{PlanckCollaboration2020} is the cosmic baryon mass fraction, 
the Lagrangian halo mass function $dn_{\rm h}/dM'$ (number density of haloes per unit mass) and the baryon fraction $f_{\rm b}$ in a halo as functions of $t$ and $M'$, which depend on the overdensity\footnote{The overdensity $\delta$ (on the scale of $3\ \rm cMpc$ corresponding to the spatial resolution of simulations) is defined such that the local baryon density follows $\rho_{\rm b}=(1+\delta)\bar{\rho}_{\rm b}$ given the cosmic average baryon density $\bar{\rho}_{\rm b}$. Similarly, we have $n_{\rm H}=(1+\delta)X_{\rm p}\bar{\rho}_{\rm b}/m_{\rm H}$, where $X_{\rm p}=0.76$ is the primordial hydrogen mass fraction, and $m_{\rm H}$ is proton mass.} $\delta$ and relative velocity $v_{\rm bc}$ between baryons and dark matter\footnote{Structure formation and collapse of gas into haloes are regulated by the relative velocity (i.e., streaming motion) between baryon and dark matter, whose density fluctuations are different at small-scales, such that the halo baryon fraction is lower than the cosmic average value $f_{\rm b,0}=\Omega_{\rm b}/\Omega_{\rm m}\simeq 0.16$ \citep{PlanckCollaboration2020} in small haloes.} in the cell \citep{Tseliakhovich2010,Barkana2011,Naoz2011,Naoz2013,Fialkov2012}, $f_{\star,\rm III}$ is the average Pop~III SFE (i.e., mass fraction of baryons in Pop~III star-forming haloes that become stars), $M_{\max}=10^{12}\ \rm M_\odot$ is the upper bound of halo mass chosen to cover most halos at $z\gtrsim 6$, and $\Delta t_{\star,\rm III}=2\ \rm Myr$ is a characteristic timescale chosen to be marginally below the lifetimes of most Pop~III stars with initial masses $M_{\star}\lesssim 300\ \rm M_\odot$. 

%Once $f_{\rm gas,\rm III}$ is known, In neutral regions of the IGM

For Pop~II stars that can form continuously in metal-enriched haloes, we instead use the following formula for both neutral and ionized regions motivated by the model in \citet{Park2019}\footnote{Following \citet{Park2019}, we estimate the star formation rate using the mass of stars ever formed divided by a characteristic timescale $t_{\star,\rm II}H(t)^{-1}$. However, we do not use the power-law halo-mass dependent SFE and the exponential suppression of star formation in small halos by SN feedback in \citet{Park2019}. Similar effects are instead modelled by our own prescriptions detailed in \citet{Fialkov2013} and \citet{Magg2022tr}.}:
\begin{align}
    \dot{\rho}_{\star,\rm II}(t)=\frac{f_{\star,\rm II}f_{\rm II}(t)}{t_{\star,\rm II}H(t)^{-1}} \int_{M_{\min}(t)}^{M_{\max}}f_{\rm sup}(M')\frac{d\rho_{\rm g}(t)}{dM'}dM'\ .\label{sfr_2}
\end{align}
Here $f_{\star,\rm II}$ is the average Pop~II SFE, $t_{\star,\rm II}=0.2$ is chosen such that $t_{\star,\rm II}H(t)^{-1}$ approximately corresponds to the characteristic dynamical time of a halo \citep{Reis2022}, $f_{\rm sup}(M')=\log(M'/M_{\min})/\log(M_{\rm atm}/M_{\min})$ is a suppression factor that captures the reduction of gas mass available for Pop~II star formation in haloes below the atomic-cooling threshold $M_{\rm atm}$ \citep[][note that $f_{\rm sup}=1$ for $M'\ge M_{\rm atm}$]{Fialkov2013}\footnote{In the original terminology of \citet{Fialkov2013}, the suppression factor $f_{\rm sup}(M')$ is introduced via a halo mass-dependent SFE $f_{\star}(M')\equiv f_{\star,\rm II}f_{\rm sup}(M')$. }, and $f_{\rm II}(t)$ is the mass fraction of star-forming haloes that host Pop~II stars, which is a function of $\delta$ and $v_{\rm bc}$, given by fits to the results in \citet[]{Magg2022tr} for the transition from Pop~III to Pop~II star formation derived from halo merger trees using the semi-analytical code \textsc{a-sloth} \citep{Hartwig2022,Hartwig2024}. In \citet{Magg2022tr}, $f_{\rm II}$ also depends on the recovery time $t_{\rm rec}$ that describes how fast a halo can restore its gas reservoir for star formation after the SN explosions of the first generation of stars. We explore the impact of $t_{\rm rec}$ on our results considering three values $t_{\rm rec}=10$, 30, and 100~Myr. In neutral regions we set the halo mass threshold as $M_{\min}=M_{\rm cool}$, while in fully ionized regions, we further consider the lower limit $M_{\rm crit}$ (above which gas can still cool to form stars under the irradiance of UV fields in ionized bubbles) imposed by photo-heating/ionization feedback \citep[][see their eq.~3]{Sobacchi2013} following \citet{Cohen2016}, such that $M_{\min}=\max(M_{\rm cool},M_{\rm crit})$.%We explore three cases with $t_{\rm rec}=10$, 30, and 100~Myr. 

To evaluate Eqs.~\ref{nion}-\ref{sfr_2}, for Pop~III stars we adopt a typical escape fraction $f_{\rm esc,III}=0.5$ based on the 1D radiation transfer calculations in \citet{Sibony2022}, which is also close to the best-fitting value $f_{\rm esc,III}=0.525$ inferred from observations with \textsc{a-sloth} \citep{Hartwig2024}. 
$\epsilon_{\rm b,III}^{\rm ion}$ is derived self-consistently by averaging the production efficiency of ionizing photons as a function of $M_\star$ (Fig.~\ref{fig:nphoton}) over the IMF (weighted by $M_\star$). We vary the average Pop~III SFE in the typical range $f_{\star,\rm III}\in [10^{-4},0.01]$ predicted by analytical models and simulations of Pop~III star formation \citep[e.g.,][]{Hirano2023,Liu2024sf}. For Pop~II stars, we fix the average SFE to $f_{\star,\rm II}=0.01$, which is chosen to reproduce the SFRD measured by galaxy surveys at $z\sim 6$ \citep[e.g.,][]{Madau2014,Finkelstein2016,Algera2023,Donnan2023,Harikane2023,Robertson2024}. %following \citet{Gessey-Jones2023}. 
{To remain connected to the convention used in previous studies, we define a phenomenological ionization efficiency parameter for Pop~II stars as $\zeta_{\rm II}\equiv \mu\epsilon_{\rm b, II}^{\rm ion}f_{\star,\rm II}f_{\rm esc,II}/(1+N_{\rm rec})$, so the Pop~II contribution in Eq.~\ref{nion} can be written as $n_{\gamma,\rm ion,II}=\zeta_{\rm II}f_{\rm coll}$, where $f_{\rm coll}\equiv f_{\rm stellar,II}/f_{\star,\rm II}$ is the cumulative mass fraction of baryons that constitute the gas reservoir for Pop~II star formation, and $\mu\sim 1.22$ is the mean molecular weight of neutral IGM.} We adopt a conservative value $\zeta_{\rm II}=4$ given $f_{\star,\rm II}=0.01$, $\epsilon_{\rm b, II}^{\rm ion}\simeq 8.5\times 10^{3}$ corresponding to a stellar population\footnote{We extrapolate the fitting formulae for the production rate of ionizing photons and lifetime as functions of initial stellar mass from \citet[see their table~6]{Schaerer2002} to derive $\epsilon_{\rm b, II}^{\rm ion}$. The rather high upper mass limit $300\ \rm M_\odot$ of the Pop~II IMF is chosen to be consistent with the Pop~III case. This choice is also supported by the metal scaling relations of high-$z$ galaxies recently observed by JWST \citep[e.g.,][]{Nakajima2023,Curti2024,Sarkar2025}, which require significant contributions of very massive ($\gtrsim 200\ \rm M_\odot$) stars to metal enrichment \citep{Liu2025}.} with a \citet{Kroupa2001} IMF from 0.1 to 300 $\rm M_\odot$ and metallicity of $Z=0.0004$, $f_{\rm esc,II}=0.15$, a typical value inferred from observations for faint galaxies (with UV magnitudes $\gtrsim -17$) at $z\sim 6-12$ \citep[][]{Chisholm2022,Mitra2023,Asthana2024}, and $N_{\rm rec}=3$ based on observations of the mean free path of ionizing photons \citep[which imply $N_{\rm rec}+1\sim 3-6$,][]{Davies2021,Davies2024}. Here $N_{\rm rec}$ is fixed for simplicity. We plan to consider the spatial and time evolution of $N_{\rm rec}$ with more self-consistent treatments of IGM clumping and attenuation in future work \citep[see, e.g.,][]{Sobacchi2014,Hassan2016,Chen2020,Mao2020,Davies2022,Cain2023,Zhu2023is}. A more comprehensive treatment of the escape fraction including the dependence on redshift (and halo/galaxy properties) is also deferred to future work \citep[e.g.,][see their fig.~1 and the references therein]{Ferrara2025}.

%Finally, in addition to $\alpha$ and $f_{\star,\rm II}$, we also explore the impact of $t_{\rm rec}$ on our results considering three values $t_{\rm rec}=10$, 30, and 100~Myr.

%The latter process is captured by the partially ionized fraction $x_{\rm e,oth}$

%Here, the Ly$\alpha$ coupling is expected to dominate 

%coupled to the background radiation temperature $T_{\gamma}$, IGM kinetic temperature $T_{\rm K}$, and 

%The reader is referred to their Sec.~3.1 for

%\subsection{X-ray heating and ionization}

%We defer a comprehensive exploration on the effect of 
%\footnote{Other than RLOF XRBs, CH Pop~III stars are more likely to produce wind-fed XRBs (that do not require very close binaries) via strong Wolf-Rayet winds \citep{Jeena2023} or the `Be-phenomenon' in which a fast-rotating star ejects materials to a decretion disk that a compact companion can accrete from \citep{{Reig2011,Rivinius2013,Liu2024}}. }

\subsection{Cosmic structure formation and simulation setup}\label{sec:setup}

As cosmic structure formation sets the stage of IGM evolution, 
the overdensity $\delta$ and baryon-dark matter relative velocity $v_{\rm bc}$ fields together with the initial IGM properties at $z_{\rm ini}=50$ serve as the foundations of our simulations through their impact on $\rho_{\rm b}$, $d\rho_{\rm g}/dM'$, $M_{\min}$, and $f_{\rm II}$. In this study, these fields are created on a $128^3$ grid cubic cells with a cell size of $3\ \rm cMpc$\footnote{The choice of the resolution is motivated by (1) the relative velocity between dark matter and gas  $v_{\rm bc}$ being coherent on scales below  $10\ \rm cMpc$, (2) for most pixels in a simulation box structure formation is either linear or mildly nonlinear at $z\gtrsim 6$ on scales above $\sim 3\ \rm cMpc$, and (3) the resolution of the SKA. Increasing resolution would not be practical from the observational point of view, and would enhance the errors caused by nonlinear growth \citep{Nikolic2024}.}. Structure formation remains marginally linear at scales larger than the cell size for $z\gtrsim 6$. Therefore, the overdensity field is evolved in a self-similar manner from the initial condition $\delta_{\rm ini}$ as $\delta(z)=D(z)\delta_{\rm ini}$, where $D(z)$ is the normalized growth factor ($D(z_{\rm ini})=1$) obtained by solving the linear perturbation equations \citep[e.g.,][]{mo2010galaxy}, $v_{\rm bc}$ simply decays with time as $v_{\rm bc}(z)\propto (1+z)$, and the halo mass function and baryon fraction are calculated analytically in each cell following \citet{Barkana2004,Fialkov2012}. The initial conditions for the fields of $\delta$ and $v_{\rm bc}$ are computed using \textsc{camb} \citep{Lewis2000,Lewis2002,Lewis2011}, while the initial IGM temperature and residual ionized fraction fields are calculated with \textsc{recfast} \citep{Seager1999,Seager2011}. 

The impact of XRBs is modelled with a dimensionless parameter $f_{\rm X}$ that relates the X-ray luminosity $L_{\rm X}$ of a galaxy to the star formation rate (SFR): $L_{\rm X}=3\times 10^{40}f_{\rm X}\ \rm erg\ s^{-1}\times SFR/(M_\odot\ yr^{-1})$. Given $f_{\rm X}$, the X-ray emissivity can be easily derived from the SFRD (Eqs.~\ref{sfr_3}~and~\ref{sfr_2}). 
Throughout this paper, we adopt a typical value $f_{\rm X,II}=1$ for Pop~II stars based on binary population synthesis predictions \citep[e.g.,][]{Fragos2013bps,Fragos2013,Liu2024} and observations in X-rays and radio \citep[e.g.,][]{Abdurashidova2022,Lehmer2022,Bevins2023,Riccio2023,Pochinda2024,Dhandha2025}. The X-ray spectrum is assumed to be a power law with a slope of $\alpha_{\rm X}=-1.5$ and lower photon energy bound of $0.1$~keV \citep{Fialkov2014xray,Fialkov2017}. The Pop~III contribution is simply ignored with $f_{\rm X,III}=0$ because our focus is the effects of UV radiation, and Pop~III XRBs are poorly understood due to the absence of direct observations and large uncertainties in binary population synthesis models of Pop~III stars \citep[see, e.g.,]{Ryu2016,Liu2021binary,Sartorio2023}. %(see the introduction). 

\begin{table}
    \centering
    \caption{Simulation parameters. The first section shows the parameters fixed throughout this work. The second section shows the parameters that are explored in physically motivated ranges.}
    \begin{tabular}{lll}
    \hline
        Symbol & Value(s)/range & Definition\\
    \hline
        $f_{\star,\rm II}$ & 0.01 & Average Pop~II star formation efficiency \\
        $t_{\star,\rm II}$ & 0.2 & Pop~II star formation timescale parameter\\
        $\zeta_{\rm II}$ & 4 & Pop~II ionization efficiency\\
        %$f_{\rm esc,II}$ & ? & Pop~II escape fraction of ionizing photons\\
        $f_{\rm X,II}$ & 1 & Pop~II galactic X-ray emission efficiency\\
        $f_{\rm X,III}$ & 0 & Pop~III galactic X-ray emission efficiency\\
        $\alpha_{\rm X}$ & -1.5 & Power-law slope of the X-ray spectrum \\
        $E_{\rm X,\min}$ & 0.1~keV & Lower energy bound of X-ray photons \\
        $N_{\rm rec}$ & 3 & Number of recombinations per baryon\\
        $R_{\max}$ & 50 cMpc & Maximum ionized bubble radius\\
        %$f_{\rm rad}$ & 0 & Galactic radio emission efficiency \\
        %$p_{\rm LW}$ & 0.75 & LW feedback delay \citep{Fialkov2013}\\
        $M_{\min}$ & $9\ \rm M_\odot$ & Minimum Pop~III mass\\
        $M_{\max}$ & $300\ \rm M_\odot$ & Maximum Pop~III mass\\
    %\hline
        $f_{\rm esc,III}$ & $0.5$ & Escape fraction of Pop~III ionizing photons\\
        $M_{z=20}$ & $5.8\times 10^{5}\ \rm M_\odot$ & Minimum mass of molecular-cooling haloes\\
        & & with Pop~III star formation at $z=20$\\
    \hline    
        $f_{\star,\rm III}$ & $10^{-4}-0.01$ & Average Pop~III star formation efficiency\\
        $\alpha$ & 0, 1, $2.35$ & Slope of Pop~III IMF: $\frac{dN}{dM_{\star}}\propto M_{\star}^{-\alpha}$\\
        $t_{\rm rec}$ &$10,30,100$ & Recovery time [Myr] from Pop~III SNe\\
        %$M_{\rm PopIII}$ & $9,300\ \rm M_\odot$ & Pop~III stellar mass for a single-mass IMF\\
        %$M_{z=20}$ & $5.8\times 10^{5}\ \rm M_\odot$ & Halo mass threshold for Pop~III star formation\\
    \hline
    \end{tabular}
    \label{tab:param}
\end{table}

As mentioned in the previous subsections, the underlying/initial fields are combined with parameterized prescriptions for star formation, stellar emission, and radiation transfer to evolve the state of the IGM. The parameters that are most relevant for this work are listed in Table~\ref{tab:param}, which are assumed to be constant throughout the simulation box and time-span for simplicity. In reality, these parameters can have non-trivial (even stochastic) time evolution, dependence on halo properties \citep[e.g.,][]{Tacchella2018,Qin2020,Cohen2017,Munoz2022,Kaur2022,Sibony2022,Harikane2023,Kim2024,Asthana2024}, and spatial fluctuations \citep[e.g.,][]{Cohen2018, Reis2022}.
%, which will be investigated in future studies (Dhandha et al. in prep., Dasgupta et al. in prep.). 
Among these parameters, we particularly focus on the average SFE $f_{\star,\rm III}$ and IMF slope $\alpha$ of Pop~III stars, and the recovery time $t_{\rm rec}$ for Pop~II star formation, which are varied in reasonable ranges motivated by the typical values from analytical models and hydrodynamic simulations of Pop~III star formation \citep[e.g.,][]{Hirano2014,Hirano2015,Hirano2023,Klessen2023,Liu2024sf} and Pop~II star formation in haloes enriched by Pop~III SNe \citep[e.g.,][]{Jeon2014,Smith2015,Chiaki2018,Chiaki2019,Latif2020,Abe2021,Magg2022,Chen2024,Hartwig2024}. The other parameters are fixed for simplicity. We also do not consider any correlations between parameters that may exist in reality\footnote{It is shown in \citet{Sibony2022} that $f_{\rm esc,III}$ increases with higher $f_{\star,\rm III}$, a more top-heavy IMF, or CHE.}. 
Throughout this paper, we assume the standard $\Lambda$CDM cosmology and use the best-fitting cosmological parameters from \citet[see the best-fit results for \textit{Planck}+WP in their table~2]{PlanckCollaboration2014}: $\Omega_{\rm m}=0.3183$, $\Omega_{\rm b}=0.0490$, $H_{0}=67.04\ \rm km\ s^{-1}\ Mpc^{-1}$, $\sigma_8=0.8347$, and $n_{\rm s}=0.9619$.
%$\Omega_{\rm m}=0.3111$, $\Omega_{\rm b}=0.0490$, $H_{0}=67.66\ \rm km\ s^{-1}\ Mpc^{-1}$, $\sigma_8=0.8102$, $n_{\rm s}=0.9665$, and $N_{\rm eff}=3.0460$. 

%Our simulations focus on the linear regime

\section{Results}\label{sec:res}
In this section, we compare the results for CH and NR Pop~III stars under different choices of $f_{\star,\rm III}$, $\alpha$, and $t_{\rm rec}$, in terms of the cosmic star formation history (Sec.~\ref{sec:sfh}), reionization (Sec.~\ref{sec:ion}), and 21-cm signal (Sec.~\ref{sec:21cm}). Focusing on the differences between NR and CH stellar evolution, we only consider two extreme cases in which  \textit{all} Pop~III stars are either NR or CH throughout their MS, although a realistic Pop~III population can be a mixture of stars in different evolution pathways \citep[see, e.g., ][]{Liu2024che}. The exact fraction of stars with CHE is determined by the initial distribution of spins, mass transfer processes in binaries, and detailed mixing mechanisms of achieving CHE \citep[see, e.g.,][]{Maeder2000,Yoon2006,Brott2011,Szecsi2015,Ghodla2022,Dall'Amico2025}, which are still poorly understood. We define $f_{\star,\rm III}=0.003$, $\alpha=1$, and $t_{\rm rec}=30$~Myr as the fiducial model. Among these three key parameters, we vary one parameter at a time while keeping the rest fixed to the fiducial choices. The other parameters governing the UV and X-ray emission from Pop~II stars are always fixed (see the first section of Table~\ref{tab:param}). To better interpret the results, we calculate the IMF-averaged emission efficiency of Pop~III stars for three representative bands, as shown in Table~\ref{tab:epsilon}.

\begin{table}
    \centering
    \caption{IMF-averaged emission efficiency (number of photons emitted per stellar baryon) of Pop~III stars in different bands. Column 1 is the Pop~III IMF slope $\alpha$ ($dN/dM_\star\propto M_\star^{-\alpha}$). Columns 2, 4, and 6 show the results for (hydrogen) ionizing photons ($13.6-100$~eV), Lyman-band photons ($10.2-13.6$~eV), and LW photons (11.2-13.6~eV), respectively, from CH Pop~III stars, while Columns 3, 5, and 7 show the results from their NR counterparts. }
    \begin{tabular}{c|ccc|ccc}
    \hline
        $\alpha$ & $\epsilon_{\rm b,III}^{\rm ion}$ & $\epsilon_{\rm b,III}^{\rm ion}$ & $\epsilon_{\rm b,III}^{\rm Ly}$ & $\epsilon_{\rm b,III}^{\rm Ly}$ & $\epsilon_{\rm b,III}^{\rm LW}$ & $\epsilon_{\rm b,III}^{\rm LW}$\\
        & (CH) & (NR) & (CH) & (NR) & (CH) & (NR)\\
        \hline 
        2.35 & 138972 & 45882 & 28251 & 21908 & 19449 & 14952 \\
        1 & 131015 & 59063 & 17376 & 21166 & 12108 & 14583 \\
        0 & 127394 & 61140 & 15224 & 21407 & 10635 & 14742 \\
%        2.35 & 169555 & 55968 & 34474 & 26728 & 23733 & 18241 \\
%        1 & 159871 & 72056 & 21213 & 25820 & 14782 & 17791 \\
%        0 & 155458 & 74605 & 18585 & 26114 & 12983 & 17984 \\
    \hline
    \end{tabular}
    \label{tab:epsilon}
\end{table}

\subsection{Star formation history}\label{sec:sfh}

\begin{figure}
    \centering
    \includegraphics[width=1\linewidth]
    {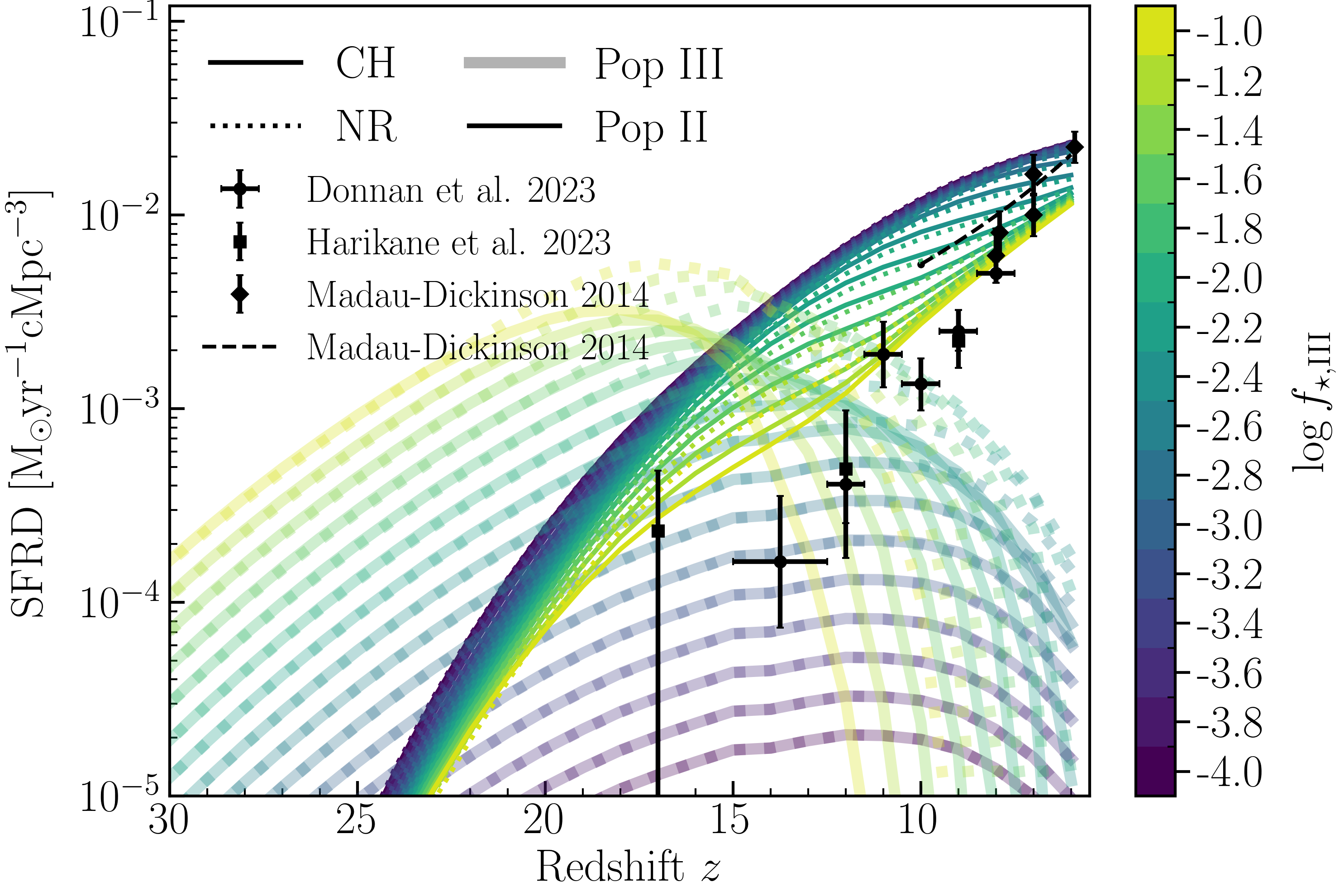}
    %{21cmSPACE_results/sfrd/SFR_fstarIII_constraints_.png}
    \vspace{-15pt}
    \caption{Cosmic SFRD for Pop~III (thick curves) and Pop~II (thin curves) stars from our simulations using the NR (dotted) and CH (solid) Pop~III stellar evolution models. The Pop~III SFE is varied in the range $f_{\star,\rm III}=10^{-4}-0.1$ with fixed $\alpha=1$ and $t_{\rm rec}=30$~Myr, where the results for higher $f_{\star,\rm III}$ are denoted by lighter colours. For comparison, the data points with $1\sigma$ errorbars show the observational results for total/Pop~II SFRD \citep{Madau2014,Donnan2023,Harikane2023} compiled by \textsc{CoReCon} \citep{Garaldi2023}. The fit ${\rm SFRD}=0.015(1+z)^{2.7}/\{1+[(1+z)/2.9]^{5.6}\}\ \rm M_{\odot}\ yr^{-1}\ Mpc^{-3}$ from \citet{Madau2014} is shown with the dashed line.}
    \label{fig:sfrd_sfe}
\end{figure}

\begin{figure}
    \centering
    \includegraphics[width=1\linewidth]{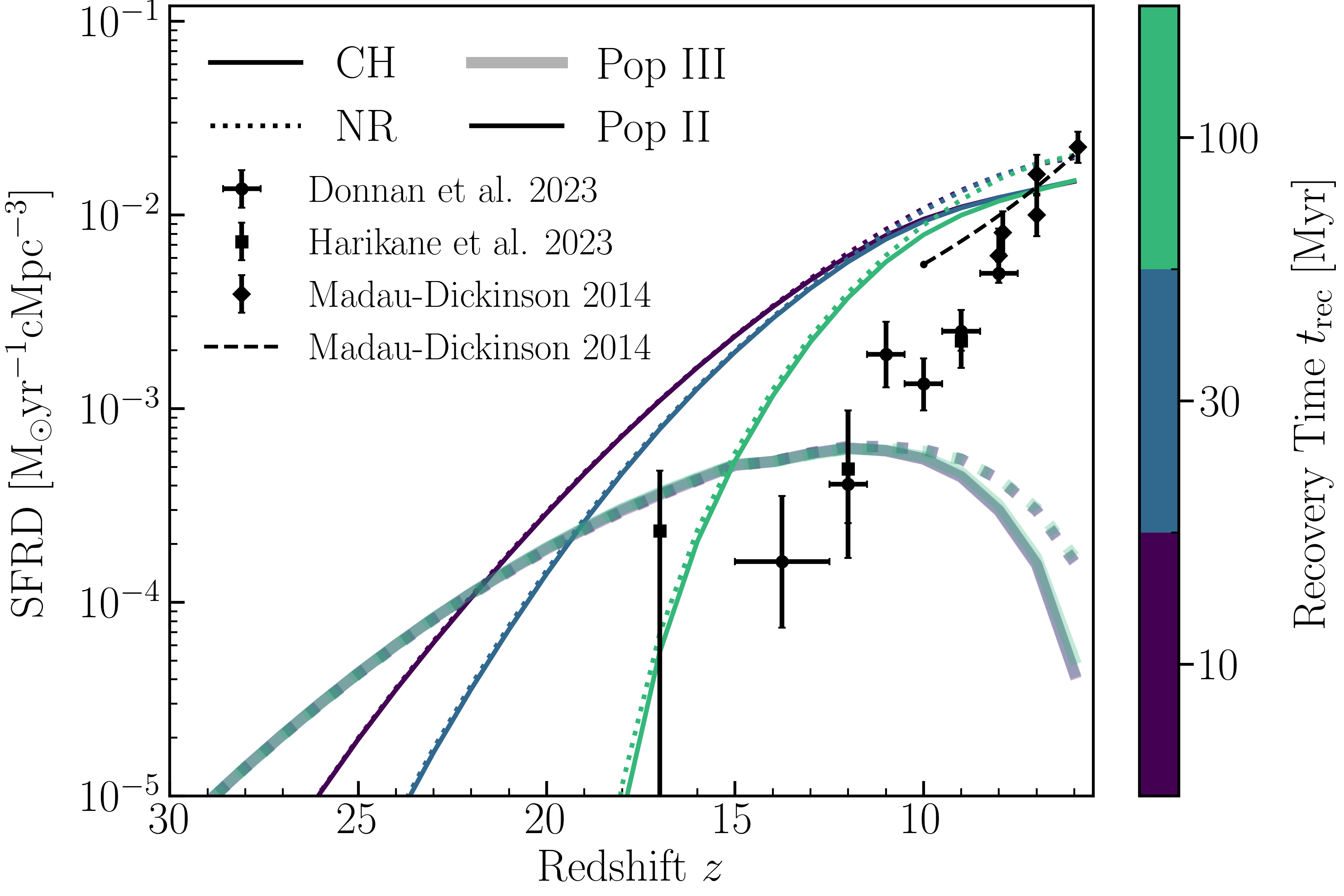}
    %{21cmSPACE_results/sfrd/SFR_trec_constraints_.png}
    \vspace{-15pt}
    \caption{Same as Fig.~\ref{fig:sfrd_sfe} but varying the recovery time in the range $t_{\rm rec}\sim 10-100$~Myr with fixed $\alpha=1$ and $f_{\star,\rm III}=0.003$. }
    \label{fig:sfrd_trec}
\end{figure}

To justify the ranges of parameters explored (Table~\ref{tab:param}), we compare the predicted SFRD with that inferred from observations \citep{Madau2014,Donnan2023,Harikane2023} in Fig.~\ref{fig:sfrd_sfe} and \ref{fig:sfrd_trec}, where $f_{\star,\rm III}$ and $t_{\rm rec}$ are varied, respectively. The observational data are obtained from the \texttt{python} package \textsc{CoReCon}\footnote{\url{https://corecon.readthedocs.io/en/latest/index.html}} \citep{Garaldi2023}. The effects of $\alpha$ on the SFRD are negligible and not shown. Cosmic star formation is always dominated by Pop~II stars at $z\lesssim 15$ where observational results have $1\sigma$ errors less than 1 dex. Interestingly, the Pop~II SFRD decreases with $f_{\star,\rm III}$. %especially for $f_{\star,\rm III}\gtrsim 0.001$ and $z\lesssim 15$ where the trend is stronger in the CH case. %Besides, the Pop~II SFRD is significantly reduced by CHE for $f_{\star,\rm III}= 0.003$ (0.01) at $z\lesssim 10$ (15). 
This is caused by the suppression of star formation by photo-heating feedback in ionized regions (where $M_{\rm min}$ is higher, see Sec.~\ref{sec:ion_model}) whose volume filling fraction is enhanced by higher $f_{\star,\rm III}$ and CHE. The effect of $f_{\star,\rm III}$ is stronger at lower $z$ for $z\sim 20-12$ given $f_{\star,\rm III}\gtrsim 0.001$ where a significant fraction ($\gtrsim 0.1$) of the IGM has already been ionized by $z\sim 12$. It is also stronger in the CH case. This is a signature of \textit{external} ionization feedback in which ionized bubbles powered by rapid star formation in over-dense regions expand to less dense regions that has not been ionized by local star formation. The correlations between galaxies and Lyman-$\alpha$/$\beta$ line transmission observed at $z\sim 5-7$ also provide evidence for such inside-out reionization \citep{Kashino2025}. 
At $z\lesssim 12$, the results from different models tend to converge due to self-regulation of star formation by feedback. The Pop~III SFRD is significantly reduced at late epochs under enhanced ionization feedback (by higher $f_{\star,\rm III}$ and/or CHE), since Pop~III star formation is forbidden in ionized regions.

For $f_{\star,\rm III}\lesssim 0.001$ and/or at $z\sim 20-30$, the reduction of Pop~II SFRD by higher $f_{\star,\rm III}$ is almost independent of $z$ but still non-negligible, which is instead driven by \textit{local} ionization feedback. In this regime, the volume fraction of ionized regions remains small. However, as early star formation is highly clustered, favouring over-dense regions, it is spatially correlated with IGM ionization, which significantly enhances local ionization feedback\footnote{Indeed, the UV radiation from one massive galaxy can significantly suppress star formation in nearby low-mass ($\lesssim 10^{9.5}\ \rm M_\odot$) haloes \citep[][]{Zhu2024}.}.  %The effect is stronger at lower $z$ for $f_{\star,\rm III}\gtrsim 0.001$ . However, for smaller 
%At higher redshifts ($z\gtrsim 15$), LW feedback drives the decrease of Pop~II SFRD with $f_{\star,\rm III}$, and the difference between the CH and NR models is very small because they have similar production rates of LW photons for $\alpha=1$. 
Meanwhile, in these over-dense regions, the transition from Pop~III to Pop~II star formation is faster and nearly complete when local ionization feedback kicks in \citep[see fig. 3 in ][]{Magg2022tr}. Therefore, the contribution of these regions to overall Pop~III star formation is always small, such that the Pop~III SFRD is insensitive to local reionization feedback, and remains almost proportional to $f_{\star,\rm III}$. 

It is shown in Fig.~\ref{fig:sfrd_trec} that the Pop~II SFRD is lower at $z\gtrsim 10$ with increasing $t_{\rm rec}$, and effect is stronger at higher $z$. The difference between $t_{\rm rec}=10$~Myr and $t_{\rm rec}=100$~Myr exceeds one order of magnitude at $z\gtrsim 17$. However, the Pop~III SFRD hardly varies, because it is only regulated by external ionization feedback at late epochs ($z\lesssim 10$ for $f_{\star,\rm III}=0.003$).  %On the other hand, the Pop~III SFRD is almost identical in the CH and NR cases (with relative differences $\lesssim 6\%$) and independent of $t_{\rm rec}$, and its normalization is almost proportional to $f_{\star,\rm III}$. This indicates that Pop~III star formation is mainly governed by $f_{\star,\rm III}$ in our model, and is relatively insensitive to the other parameters that indirectly regulate Pop~III star formation via ionization and LW feedback. %(whose effects tend to cancel out each other\footnote{When reionization is accelerated, formation of Pop~II stars will be reduced, leading to a weaker LW background. As a result, Pop~III star formation will be enhanced in neutral regions, which compensates the reduction by the increased volume of ionized regions.}). 
%, because ionized bubbles in over-dense regions

%\citep[e.g.,][]{Johnson2013,Smith2015,Xu2016,Sarmento2017,Sarmento2018,El-Badry2018,Liu2019,Liu2020sim,Liu2020,Liu2022,Skinner2020,Kulkarni2021,Schauer2019vbc,Schauer2021,Kulkarni2022,Yajima2022,Yajima2023,Kiyuna2023,Garcia2023,Venditti2023,Incatasciato2023,CorreaMagnus2024,Lenoble2024,Sugimura2024,Smith2024} and semi-analytical models \citep[e.g.,][]{Manrique2015,Salvador-Sole2017,Griffen2018,Dayal2020,Visbal2020,Li2021,Lupi2021,Hartwig2022,Trinca2022,Trinca2024,Hegde2023,Nebrin2023,Bovill2024,Ventura2024,Feathers2024}

\begin{figure*}
    \centering
    \includegraphics[width=1\linewidth]{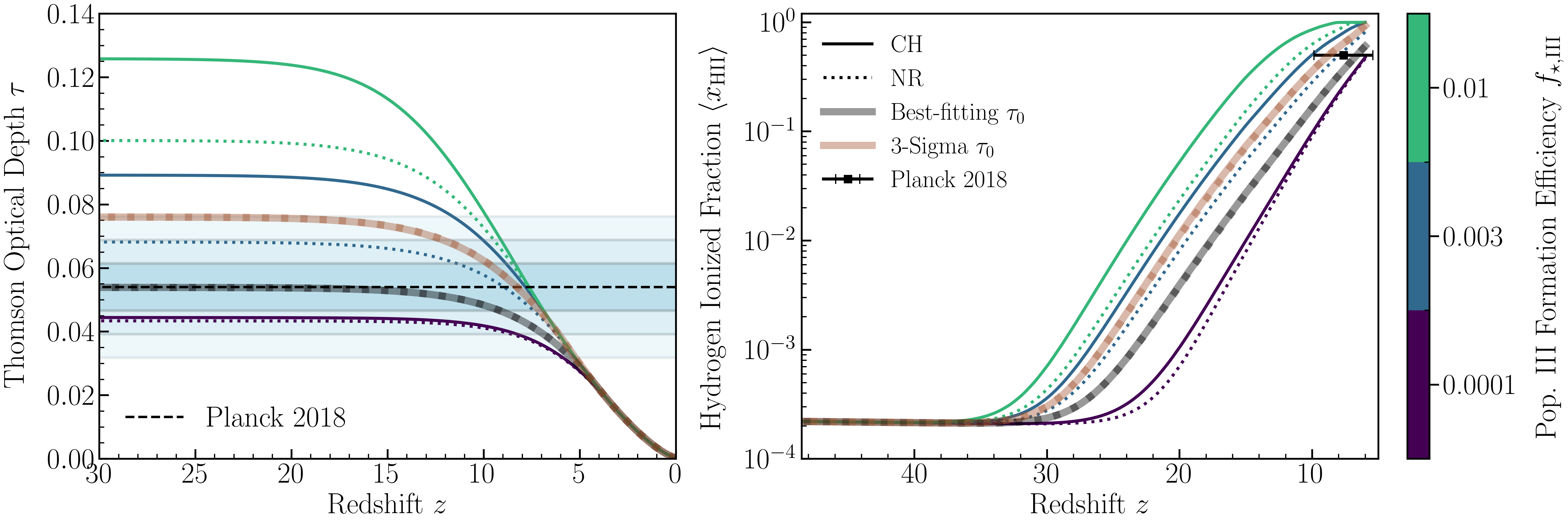}
    %{21cmSPACE_results/reion/reion_fstarIII_crit.png}
    \vspace{-15pt}
    \caption{Ionization history in terms of the optical depth (left) and volume-averaged ionized fraction (right). The dotted (solid) curves show the results for the NR (CH) Pop~III stellar evolution models, where the Pop~III SFE is varied in the range $f_{\star,\rm III}=10^{-4}-0.01$ (as marked on the colorbar) with fixed $\alpha=1$ and $t_{\rm rec}=30$~Myr. Here, the results for higher $f_{\star,\rm III}$ are denoted by lighter colours. In the left panel, the measurement of $\tau_{0}\equiv\tau(z\rightarrow 1100)$ by Planck $\tau_{0}=0.0544\pm0.0073$ \citep{PlanckCollaboration2020} is shown with the dashed line and shaded regions for $1-3\sigma$ errors. In the right panel, the errorbar denotes the $3\sigma$ confidence interval of the redshift of instantaneous reionization $z_{\rm re}= 7.67\pm 0.73$ (when $\langle x_{\rm HII}\rangle=0.5$) inferred from observations \citep{PlanckCollaboration2020}. We also find the $f_{\star,\rm III}$ values required to produce the best-fitting value $\tau_0=0.0544$ ($f_{\star,\rm III}\sim 5.7\times 10^{-4}$ and $1.3\times 10^{-3}$ for CH and NR stars) and the $3\sigma$ upper limit $\tau_0=0.0763$ ($f_{\star,\rm III}=1.8\times 10^{-3}$ and $4.2\times 10^{-3}$ for CH and NR stars), as shown by the lower and upper thick curves, respectively. 
    }
    \label{fig:reion_sfe}
\end{figure*}

Our results for the total/Pop~II SFRD are generally consistent with observations at $z\lesssim 7$, but the SFRD at higher redshifts is over predicted by up to $\sim 1$~dex around $z\sim 14$. Since $f_{\star,\rm II}$ is constant in our model, to match the observations at higher redshifts, one must increase $t_{\rm rec}$ and/or decrease $f_{\star,\rm II}$ at the price of under predicting the SFRD at lower redshifts where observational results are less uncertain. This implies that $t_{\rm rec}\gtrsim 100$~Myr is favoured and/or our star formation model needs to be improved to fully reproduce observations, which is beyond the scope of this paper \citep[see][]{Dhandha2025}\footnote{Note that the SFRD inferred from JWST observations \citep{Donnan2023,Harikane2023} is derived by integrating the UV luminosity function up to the magnitude of $M_{\rm UV}=-17$, which corresponds to $\rm SFR\sim 0.3\ \rm M_\odot\ yr^{-1}$ according to the canonical UV-SFR conversion coefficient \citep{Madau2014}. However, the SFRD predicted by our simulations covers all galaxies, including those with $\rm SFR\lesssim 0.3\ \rm M_\odot\ yr^{-1}$. The contributions of such smaller galaxies can be large depending on the shape of the UV luminosity function, which my also explain the discrepancy. More direct constraints on star-formation parameters can be obtained with the UV luminosity function itself rather than its integration \citep{Dhandha2025,ZviKatz2025}}. Better agreements with observations can be achieved if the SFE decreases with halo mass in low-mass haloes, which is a natural consequence of stellar feedback and a common prescription in galaxy formation models \citep[e.g.,][]{Tacchella2018,Behroozi2019,Shen2023}. In our case with an optimistic \textit{constant} value $f_{\star,\rm II}=0.01$ and the simulation-motivated choices of $t_{\rm rec}\sim 10-100$~Myr, the contribution by Pop~II stars is likely overestimated at $z\gtrsim 7$. Therefore, the effects of Pop~III stars (with respect to those of Pop~II stars) in our simulations should be regarded as conservative estimates. Nevertheless, our Pop~III star formation histories are consistent with the predictions from previous analytical models and cosmological simulations \citep[e.g.,][]{Johnson2013,deSouza2013,deSouza2014,Smith2015,Xu2016,Mebane2018,Sarmento2017,Sarmento2018,Sarmento2019,Jaacks2018,Jaacks2019,Liu2020sim,Liu2020,Dayal2020,Visbal2020,Skinner2020,Hartwig2022,Munoz2022,Venditti2023,Ventura2024}. In particular, as shown below, we obtain a similar range of the cumulative stellar mass density (CSMD) of Pop~III stars (ever formed) $\rho_{\star,\rm III}\sim 10^{4}-10^{6}\ \rm M_\odot\ Mpc^{-3}$ as that covered by the above literature.

\subsection{Ionization history}\label{sec:ion}

% (varying $f_{\star,\rm III}$)

For simplicity, we characterize the progress of reionization with the global ionization history captured by the volume-averaged (hydrogen) ionized fraction $\langle x_{\rm HII}(z)\rangle\equiv 1-\langle x_{\rm HI}(z)\rangle$ and the optical depth for Thomson scattering \citep{Robertson2013,Hartwig2015}:
\begin{align}
    \tau(z)=c\sigma_{\rm T}\int_{0}^{z}f_{\rm e}\langle x_{\rm HII}(z')\rangle \bar{n}_{\rm H}(z')\left|\frac{dt}{dz'}\right|dz'\ \label{tau}
\end{align}
given the cross-section of Thomson scattering $\sigma_{\rm T}$, the cosmic average physical hydrogen number density $\bar{n}_{\rm H}(z')\equiv X_{\rm p}\bar{\rho}_{\rm b}(z')/m_{\rm H}$, and the number of free electrons per hydrogen nucleus $f_{\rm e}=1+Y_{\rm p}/(4X_{\rm p})$, 
%\begin{align}
%    f_{\rm e}=\begin{cases}
%        1+Y_{\rm p}/(2X_{\rm p})\ ,\quad z\le 4\ ,\\
%        1+Y_{\rm p}/(4X_{\rm p})\ ,\quad z> 4\ ,
%    \end{cases}\label{fe}
%\end{align}
where $X_{\rm p}=0.76$ and $Y_{\rm p}=0.24$ are the primordial mass fractions of hydrogen and helium. Since our simulations stop at $z=6$, we extrapolate $\langle x_{\rm HI}(z)\rangle$ down to $z=0$ as a linear function of $\log(1+z)$ (bounded by the lower limit 0). The contribution of helium double ionization is neglected for simplicity, which is expected to be small since helium double ionization can only increase $f_{\rm e}$ by up to $\sim 7\%$ reaching $f_{\rm e}=1+Y_{\rm p}/(2X_{\rm p})$ at $z\lesssim 3$. 
%here we assume that the IGM is fully ionized ($\langle x_{\rm HI}\rangle=0$) at $z<6$ for hydrogen and helium single ionization, while helium double ionization happens instantaneously at $z=4$. 
The value $\tau_{0}=0.0544\pm0.0073$ measured from the CMB \citep[][]{PlanckCollaboration2020} is compared with our predictions for $\tau_0\equiv \tau(z\rightarrow 1100)$ to put constraints on Pop~III properties.

We find that the ionization history is mostly sensitive to $f_{\star,\rm III}$ and the stellar evolution model adopted, while the effects of $\alpha$ and $t_{\rm rec}$ are minor (with small changes of $\Delta\tau_0\lesssim 0.005$). Fig.~\ref{fig:reion_sfe} shows the results for the fiducial Pop~III IMF ($\alpha=1$) and recovery time ($t_{\rm rec}=30$~Myr) with varying Pop~III SFE $f_{\star,\rm III}\sim 10^{-4}-0.01$. The results for varying $\alpha$ and $t_{\rm rec}$ are shown in Appendix~\ref{apdx:reion}. Reionization is accelerated by CHE and higher $f_{\star,\rm III}$: For $f_{\star,\rm III}\gtrsim 0.001$, reionization generally proceeds faster by $\Delta z\sim 2$ in the CH models compared with their NR counterparts, such that $\langle x_{\rm HII}\rangle$ is enhanced by a factor of $\sim 2$ at $z\sim 30-10$. The impact of CHE is weaker for lower $f_{\star,\rm III}$ and redshifts where Pop~II stars dominate the ionization budget. In the extreme case with $f_{\star,\rm III}=0.01$, both CH and NR models are ruled out by observations under our assumptions on Pop~II star formation and the relevant contribution to $\tau_0$ (see below). We find that $f_{\star,\rm III}\simeq 5.74\times 10^{-4}$ and $1.28\times 10^{-3}$ are required for the CH and NR models to reproduce the best-fitting value of optical depth from \citet{PlanckCollaboration2020} $\tau_{0}=0.0544$, while $f_{\star,\rm III}\simeq 1.85\times 10^{-3}$ ($4.17\times 10^{-3}$) leads to $\tau_0=0.0763$, the 3$\sigma$ upper limit of $\tau_0$ measured by \citet{PlanckCollaboration2020}, in the CH (NR) model. This indicates that the ionization power of Pop~III stars is boosted by a factor of $\simeq 2$ with CHE for a log-flat IMF in the range of $M_\star\in[9-300]\ \rm M_\odot$. For all the models that satisfy $\tau_0<0.0763$, the Pop~III contribution to $\tau$ is mostly achieved at $z\sim 8-15$, and the contribution from the earlier epoch ($z\gtrsim 15$) remains small ($\Delta\tau\lesssim 0.01$), below the upper limit of 0.02 from \citet{PlanckCollaboration2020}, consistent with the findings in \citet{Munoz2022}.

\begin{figure}
    \centering
    \includegraphics[width=1\linewidth]{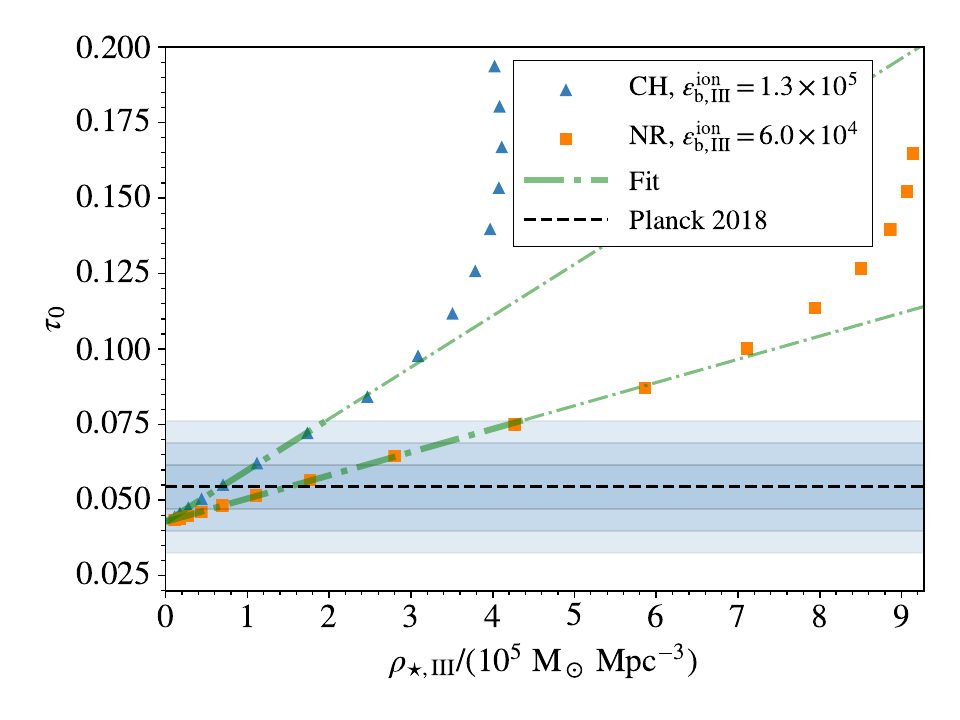}%{tau_CSMD.pdf}
    \vspace{-15pt}
    \caption{Relation between $\tau_0$ and the CSMD of Pop~III stars. The simulation results for the fiducial CH and NR models (with $\alpha=1$) are shown with the triangles and squares. The thick dash-dotted lines show the general linear fit (Eq.~\ref{tau_fit}) valid for $\tau_0<0.0763$ with relative errors $\lesssim 1\%$, whose extrapolation is shown by the thin dash-dotted lines. The Planck measurement of $\tau_{0}$ \citep{PlanckCollaboration2020} is shown with the dashed line and shaded regions for $1-3\sigma$ errors.}%{\color{orange}The CH value of $\epsilon_{\rm b, III}^{\rm ion}$ here is higher than that in Table~\ref{tab:epsilon} because of a confusion in the code regarding the definition of $\zeta$ and $\epsilon_{\rm b}^{\rm ion}$, which caused systematic underestimations of the ionizing power of NR stars. We have clarified the definitions in Eq.~\ref{epsilon_b} and the end of Sec.~\ref{sec:ion_model}, such that Table~\ref{tab:epsilon} provides the updated values. The results for all NR models have yet to be updated (including this figure).}}
    \label{fig:tau_csmd}
\end{figure}

To further quantify the contribution of Pop~III stars to reionization, we derive the relation between $\tau_0$ and the Pop~III CSMD $\rho_{\star,\rm III}$ through a series of simulations, as shown in see Fig.~\ref{fig:tau_csmd}. For $\tau_0<0.0763$, we obtain a general linear relation (covering both CH and NR models): %{\color{orange}We need to update the slope of this fit.}
\begin{align}
    &\tau_0 \simeq \tau_{\rm II}\notag\\
    &+ 0.0130\left(\frac{f_{\rm esc,III}}{0.5}\right)\left(\frac{4}{1+N_{\rm rec}}\right)\left(\frac{\epsilon_{\rm b,III}^{\rm ion}}{10^{5}}\right)\left(\frac{\rho_{\star,\rm III}}{10^5\ \rm M_\odot\ Mpc^{-3}}\right)\ ,\label{tau_fit}
\end{align}
which is valid with relative errors in $\tau_0$ less than $\sim 1\%$. Here, $\tau_{\rm II}$ is the contribution to $\tau_0$ by Pop II stars (corresponding to $f_{\star,\rm III}\rightarrow 0$), and we have $\tau_{\rm II}\simeq 0.043$ for our choice of fiducial parameters $\zeta_{\rm II}=4$ and $\tau_{\rm rec}=30$~Myr. This linear relation (Eq.~\ref{tau_fit}) provides similar constraints on Pop~III star formation from reionization compared with the earlier results $\rho_{\star,\rm III}\lesssim 10^{4}-10^{6}\ \rm M_\odot\ Mpc^{-3}$ in \citet{Visbal2015} and \citet{Inayoshi2016}. For instance, $\tau_{0}<0.0763$ requires $\rho_{\star,\rm III}\lesssim 2.0\ (4.4)\times 10^{5}\ \rm M_\odot\ Mpc^{-3}$ for CH (NR) Pop~III stars with $\alpha=1$ and $\epsilon_{\rm b,III}^{\rm ion}=1.3\times 10^{5}\ (6.0\times 10^{4})$ given $f_{\rm esc,III}=0.5$ and $\tau_{\rm II}\simeq 0.043$. The increase of $\tau_0$ with $\rho_{\star,\rm III}$ remains quasi-linear for $\tau_0\lesssim 0.1$ when the IGM is not fully ionized at $z\gtrsim 10$, which covers all NR models with $f_{\star,\rm III}\le 0.01$. %in the NR case due to the completing effects of saturation (with less efficient increase of $\tau_0$) and ionization feedback (which reduceds Pop~III SFRD) as $\tau_0$. 
For higher $\tau_0$, the $\tau_0$-$\rho_{\star,\rm III}$ relation becomes superlinear driven by (external) ionization feedback, indicating strong self-regulation %\footnote{Self-regulation of Pop~III star formation starts to be important already at $\tau_0\sim 0.08$, corresponding to $f_{\star,\rm III}\sim 0.002\ (0.005)$ in the CH (NR) case, beyond which $\rho_{\star,\rm III}$ is no longer approximately proportional to $f_{\star,\rm III}$. The $\tau_0$-$\rho_{\star,\rm III}$ relation remains linear for $\tau_0\sim 0.08-0.1$ due to the competing effects of saturation (with less efficient increase of $\tau_0$ with $f_{\star,\rm III}$) and ionization feedback (which reduces Pop~III SFRD).} 
of Pop~III star formation. While increasing $f_{\star,\rm III}$ further beyond 0.01 (up to 0.1), we find an upper limit of $\rho_{\star,\rm III,\max}\sim 4\ (9)\times 10^{5}\ \rm M_\odot\ Mpc^{-3}$ at $f_{\star,\rm III}\gtrsim 0.03\ (0.1)$ in the CH (NR) case caused by self-regulation. 

These constraints are sensitive to $f_{\rm esc,III}$, $N_{\rm rec}$, and $\tau_{\rm II}$. The Pop~III escape fraction $f_{\rm esc,III}$ can deviate from our fiducial choice by up to a factor of 2 according to the predictions $f_{\rm esc,III}\sim 0.2-0.8$ from \citet{Sibony2022}. The recombination number $N_{\rm rec}$ is observationally unconstrained at $z\gtrsim 10$ where Pop~III stars are important. The value adopted here is based on observations at the end of reionization \citep[$z\sim 5-6$,][]{Davies2021,Davies2024}. It is typically found in simulations that $N_{\rm rec}$ decreases towards higher $z$ down to $N_{\rm rec}\sim 1$ at $z\gtrsim 10$ \citep[e.g.,][]{Sobacchi2014,Hassan2016,Chen2020,Mao2020,Davies2022,Cain2023,Zhu2023is}. The Pop~II contribution $\tau_{\rm II}$ is determined by the ionization efficiency $\zeta_{\rm II}$ and $t_{\rm rec}$. It is shown in Appendix~\ref{apdx:reion} that varying $t_{\rm rec}$ only changes $\tau_{\rm II}$ by up to a few percent. However, there are larger uncertainties in $\zeta_{\rm II}$ \citep[e.g.,][]{Davies2021,Davies2024,Chisholm2022,Davies2022,Mitra2023,Asthana2024} that can modify $\tau_{\rm II}$ by a factor of a few. In particular, the total SFRD at $z\sim 8-15$ may be overestimated in our simulations under constant Pop~II SFE $f_{\star,\rm II}=0.01$. Smaller values of $f_{\star,\rm II}$ (at higher $z$) may be required to reproduce the observed SFRD, leading to smaller $\zeta_{\rm II}$ and $\tau_{\rm II}$. Considering such uncertainties, it is challenging to probe Pop~III properties by reionization alone. Generally speaking, when $\tau_{\rm II}$ is larger, any constraints on Pop~III stars from reionization will be weakened. Combining observations of reionization and the 21-cm signal can hopefully provide stronger constraints, as discussed below. Such joint constraints can be stronger with increasing $f_{\rm esc,III}$ or decreasing $N_{\rm rec}$.

\subsection{21-cm signal}\label{sec:21cm}

To be consistent with the primary interest of current and upcoming radio experiments \citep[e.g.,][]{Koopmans2015,Bowman2018,Mertens2021,Mertens2025,deLeraAcedo2022,Singh2022,HERACollaboration2023,Monsalve2023,Zhao2024}, we focus on the sky-average global signal $\langle T_{21}(z)\rangle$ and power spectrum $\Delta^{2}$ of the spatial variations of $T_{21}$: %\citep[e.g.,][]{Gessey-Jones2023}:
\begin{align}
    \langle\tilde{T}_{21}({\bm k},z)\tilde{T}_{21}^{*}(\bm{k}',z)\rangle=(2\pi)^{3}\delta^{D}({\bm{k}-\bm{k}'})\frac{2\pi^{2}}{k^{3}}\Delta^{2}(k,z)\ ,\label{powspec}
\end{align}
where $\bm k$ is the comoving wave vector and $\delta^{D}$ is the Dirac delta function. Redshift-space distortions are included in the calculations of $\langle T_{21}(z)\rangle$ and $\Delta^{2}(k,z)$.

\begin{figure*}
    \centering
    \includegraphics[width=1\linewidth]{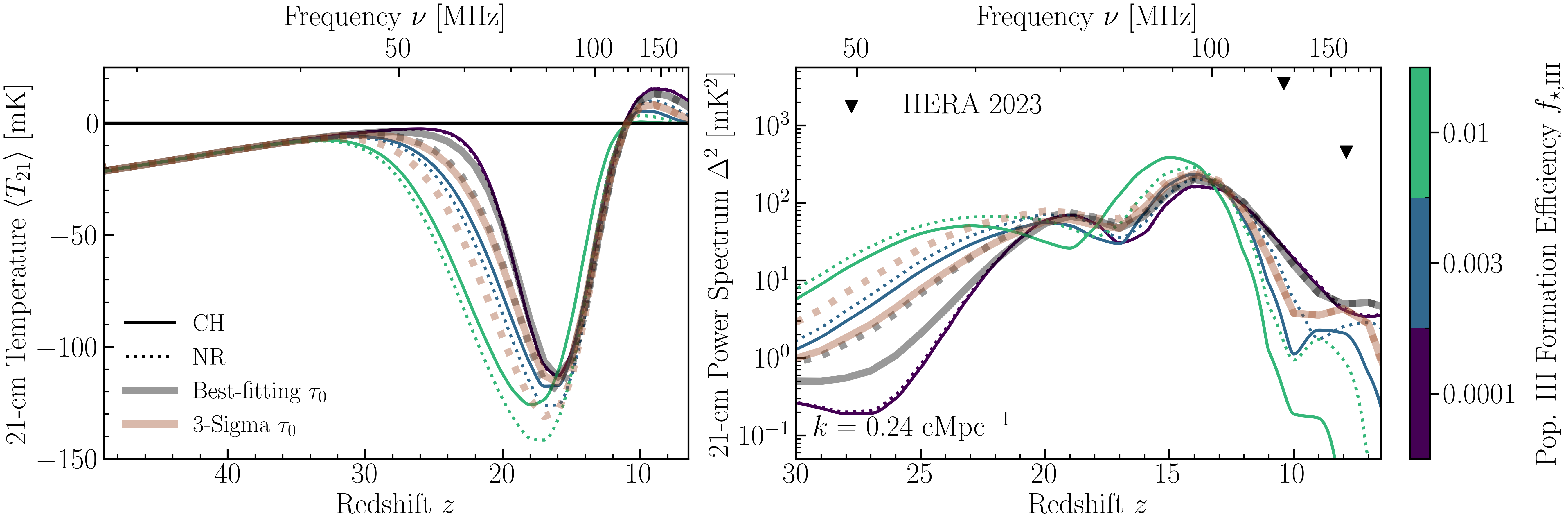}
    %{21cmSPACE_results/21cm/21cm_fstarIII_crit.png}
    \vspace{-15pt}
    \caption{Redshift evolution of the 21-cm global signal (left) and power spectrum (right) with varying $f_{\star,\rm III}\sim 10^{-4}-0.01$ (marked on the colorbar) and fixed $\alpha=1$ and $t_{\rm rec}=30$~Myr. As in Fig.~\ref{fig:reion_sfe}, the dotted (solid) curves show the results for the NR (CH) Pop~III stellar evolution models, and the results for higher $f_{\star,\rm III}$ are denoted by lighter colours. The gray and brown thick curves show the models with $f_{\star,\rm III}$ calibrated to the best-fitting value and $3\sigma$ upper limit of $\tau_0$ in observations \citep{PlanckCollaboration2020}, respectively. In the right panel, we show the power at a characteristic scale $k=0.24\rm\ cMpc^{-1}$, compared with the observational constraints ($2\sigma$ upper limits) from \citet[]{HERACollaboration2023}, which are denoted as triangles. }%, to be compared with the observational constraints from EDGES \citep{Bowman2018} and HERA \citep{HERACollaboration2023}.}
    \label{fig:21cm_sfe}
\end{figure*}

\begin{figure*}
    \centering
    \includegraphics[width=1\linewidth]{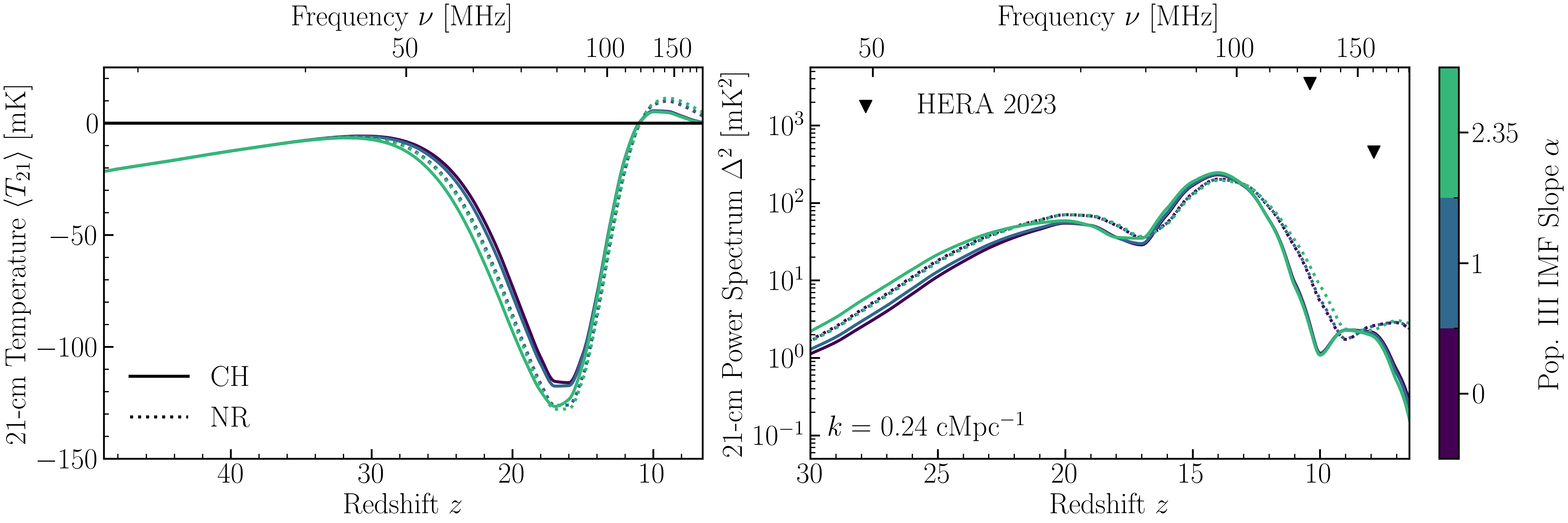}
    %{21cmSPACE_results/21cm/21cm_alpha.png}
    \vspace{-15pt}
    \caption{The 21-cm signal. Same as Fig.~\ref{fig:21cm_sfe} but varying the Pop~III IMF slope $\alpha$ (0, 1, and 2.35) with fixed SFE $f_{\star,\rm III}=0.003$ and recovery time $t_{\rm rec}=30$~Myr.}
    \label{fig:21cm_alpha}
\end{figure*}

\begin{figure*}
    \centering
    \includegraphics[width=1\linewidth]{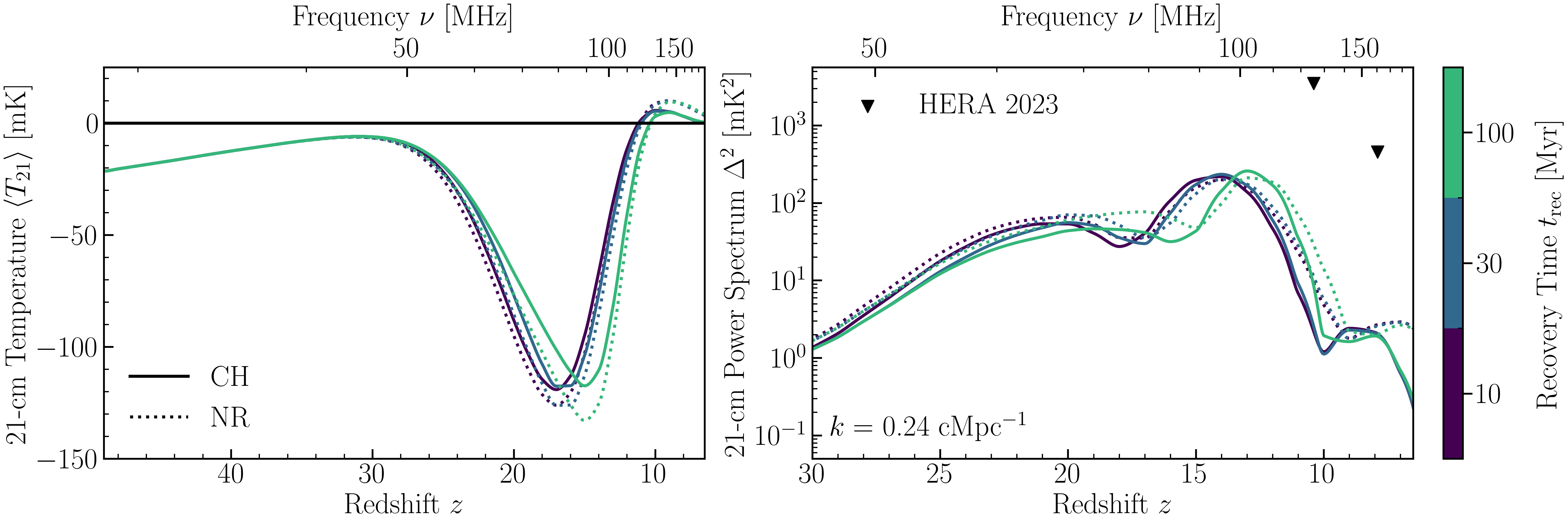}
    %{21cmSPACE_results/21cm/21cm_trec.png}
    \vspace{-15pt}
    \caption{The 21-cm signal. Same as Fig.~\ref{fig:21cm_sfe} but varying the recovery time $t_{\rm rec}$ (10, 30, and 100~Myr) with fixed Pop~III SFE $f_{\star,\rm III}=0.003$ and IMF slope $\alpha=1$.}
    \label{fig:21cm_trec}
\end{figure*}

Figure~\ref{fig:21cm_sfe} compares the 21-cm predictions from CH (solid) and NR (dotted) models with different values of Pop~III SFE $f_{\star,\rm III}\sim 10^{-4}-0.01$ with fixed $\alpha=1$ and $t_{\rm rec}=30$~Myr. The global signal (left panel) is characterized by an absorption ($\langle T_{21}\rangle<0$) trough around $z\sim 12-30$ followed by a transition to emission and the subsequent decay at $z\lesssim 8$. %The depth and timing of the absorption trough are sensitive to the stellar UV and X-ray radiation. 
The rise of absorption (starting from $z\sim 32-24$ and reaching the peak at $z\sim 17$) marks the epoch of coupling (EoC), when the spin temperature $T_{\rm S}$ of neutral hydrogen is driven towards the gas kinetic temperature $T_{\rm K}$ by the WF effect. Then in the epoch of heating (EoH), $\langle T_{21}\rangle$ increases along with $T_{\rm K}$ due to the heating of IGM by X-rays, until it becomes positive around $z\sim 11$. Thereafter, $\langle T_{21}\rangle$ converges to zero at lower redshifts as the IGM is increasingly ionized in the EoR. 
%The transition from absorption to emission is mainly governed by heating from X-rays
As the Ly$\alpha$ radiation fields become stronger with higher $f_{\star,\rm III}$, facilitating the WF effect during EoC, the absorption trough is deepened and shifted to higher redshifts. The maximum absorption signal varies between $\langle T_{21}\rangle\sim -120$~mK and $\langle T_{21}\rangle\sim -150$~mK, while the timing of peak absorption is relatively stable at $z\sim 16-18$, consistent with the timing of the tentative signal detected by the Experiment to Detect the Global EoR Signature (EDGES) \footnote{Whether this signal has an astrophysical origin is still in debate \citep[e.g.,][]{Hills2018,Bradley2019,Singh2019,Singh2022,Sims2020}. Therefore, we do not show the EDGES results on our plots.} \citep{Bowman2018}. Compared with NR stars, the absorption is shallower for CH stars given $\alpha=1$, and the difference increases with $f_{\star,\rm III}$, reaching $\sim 20$~mK at $f_{\star,\rm III}=0.01$. The main reason is that the Lyman-band emission is slightly weaker under CHE (Table~\ref{tab:epsilon}), because massive ($M_\star\gtrsim 50\ \rm M_\odot$) CH stars are too hot (see Figs.~\ref{fig:spec} and \ref{fig:nphoton}). %This effect becomes stronger when the IMF is more top-heavy (with smaller $\alpha$, see below).
Considering the constraints from reionization discussed above, we compare the CH and NR models calibrated to the best-fitting value and 3-sigma upper limit of the optical depth for Thomson scattering $\tau_0$ from \citep[see the thick curves in Figs.~\ref{fig:reion_sfe} and \ref{fig:21cm_sfe}]{PlanckCollaboration2020}, finding that the peak absorption is weaker in the CH case by $\sim 5$~mK (4\%) and 15~mK (11\%) for $\tau_0=0.0544$ (best-fitting) and $\tau_0=0.0763$ ($3\sigma$ upper limit), respectively. 
The evolution during EoH is mainly regulated by X-rays, which in our case are only produced by Pop~II stars, whose formation history has a relatively weak, indirect dependence on $f_{\star,\rm III}$ (see Fig.~\ref{fig:sfrd_sfe}). This explains why the late stage of the transition to emission ($z\lesssim 14$) is almost identical among all cases, except for the CH model with $f_{\star,\rm III}=0.01$ which shows a faster transition. In this special case already ruled out by the Planck measurement of $\tau_0$ (where even Pop~III stars alone can produce $\tau_0\sim 0.082$ above the Planck $3\sigma$ upper limit $\tau_0=0.0763$), reionization is significantly accelerated by CHE, such that the volume-averaged ionized fraction of the IGM is already high ($\gtrsim 40$\%) at $z\lesssim 14$ (Fig.~\ref{fig:reion_sfe}), reducing the overall strength of the signal (for both absorption and emission). For similar reasons, after the transition, the emission signal decays faster in the CH models and/or for higher $f_{\star,\rm III}$ during EoR.

The 21-cm power spectrum (right panel of Fig.~\ref{fig:21cm_sfe}) is sensitive to the spatial fluctuations of the Ly$\alpha$ radiation fields, IGM temperature, and ionized/neutral fractions, which dominate the signal during EoC, EoH, and EoR, respectively. Here we focus on the power at a characteristic scale $k=0.24\ \rm cMpc^{-1}$ where \citet[see their fig.~30]{HERACollaboration2023} places the strongest observational constraints on the upper limits of $\Delta^{2}$. The power spectra as a function of $k$ at fixed redshifts are shown in Appendix~\ref{apdx:21cm} for $k\sim 0.03-1\ h\ \rm cMpc^{-1}$. 
At such a relatively large scale, there are three peaks/plateaus in the redshift evolution of 21-cm power spectrum, generally corresponding to the three epochs discussed above. The troughs between them arise from negative cross correlations between the aforementioned fields. 
The first peak (at $z\sim 19-23$) becomes broader with significantly stronger (up to two orders of magnitude) power at $z\gtrsim 24$ (EoC) when $f_{\star,\rm III}$ increases, which accelerates Pop~III star formation and enhance the Ly$\alpha$ fields at larger scales during EoC. Nevertheless, the peak power only changes moderately, remaining around $\Delta^{2}\sim 50-80\ \rm mK^{2}$. The peak power is slightly weaker in the CH models compared with the NR case, as expected from the reduction of Lyman-band emission by CHE. 
When $\tau_0$ is controlled (thick curves), the power from CH stars is lower than that from NR stars by a factor of $\sim 2-3$ during EoC ($z\sim 24-30$) due to weaker Ly$\alpha$ coupling with reduced $f_{\star,\rm III}$. 
The second (strongest) peak around $z\sim 14$ with $\Delta^{2}\sim 200\ \rm mK^{2}$ shows very small variations for the models that are consistent with the observational constraints from reionization (i.e., $f_{\star,\rm III}\lesssim 0.002$ and $f_{\star,\rm III}\lesssim 0.004$ for CH and NR stars, respectively), similar to the behavior of the global signal during EoH. Again, the reason is that EoH is dominated by X-ray heating from Pop~II stars and these models have almost identical Pop~II star formation histories (Fig.~\ref{fig:sfrd_sfe}). However, for models with larger $f_{\star,\rm III}$ and accelerated IGM ionization, like in the extreme case of $f_{\star,\rm III}=0.01$, the peak is enhanced and shifted to higher redshifts, reaching up to $\Delta^{2}\sim 500\ \rm mK^{2}$ at $z=15$ for CH stars. This is caused by the suppression of Pop~II star formation in ionized regions, i.e., external ionization feedback, which reduces X-ray heating around ionized bubbles, leading to less negative contributions from the cross power between X-ray heating and neutral hydrogen density. %, i.e., correlations between emission ($T_{21}>0$) and absorption ($T_{21}<0$) regions. 
Finally, when IGM ionization is facilitated either by CHE or increasing $f_{\star,\rm III}$, the power during EoR ($z\lesssim 12$) decreases dramatically, and the third peak/plateau (at $z\lesssim 9$) is narrower and shifted to higher redshifts, which covers a abroad range $\Delta^{2}\sim 0.2-7\ \rm mK^{2}$. 

Next, we explore the effects of $\alpha$ and $t_{\rm rec}$ while fixing Pop~III SFE to the fiducial value $f_{\star,\rm III}=0.003$, as shown in Figs.~\ref{fig:21cm_alpha} (for $\alpha\sim 0-2.35$ and $t_{\rm rec}=30$~Myr) and \ref{fig:21cm_trec} (for $t_{\rm rec}\sim 10-100$~Myr and $\alpha=1$), respectively. The effects are generally smaller with $\alpha$ compared to those of $f_{\star,\rm III}$, especially for NR stars. This is reasonable because the lower mass bound of our IMFs is fixed to a relatively high value ($9\ \rm M_\odot$), above which the production efficiency $\epsilon_{\rm b}(M_\star)$ of UV photons is not very sensitive to stellar mass (Fig.~\ref{fig:nphoton}). Besides, it is shown in \citet{Gessey-Jones2022} that the 21-cm signal varies modestly across a broader range of IMFs for NR Pop~III stars if one only considers the impact of IMF on Lyman band radiation. The variation can be significantly boosted when the impact of Pop~III IMF on XRBs is considered \citep[]{Sartorio2023,Gessey-Jones2025}, which we leave to future work. CHE weakens the global absorption signal and generally reduces the power during EoC ($z\gtrsim 17$) for $\alpha\lesssim 1$, while the opposite happens for $\alpha=2.35$. In the former (latter) case, the IMF-averaged Lyman-band emission from CH stars is reduced (enhanced) compared with NR stars, as shown Table~\ref{tab:epsilon}. Due to this trend reversal, the CH models show a much stronger dependence on $\alpha$, where the peak global absorption signal varies by $\sim 10$~mK for $\alpha\sim 0-2.35$. Yet in the CH model with $\alpha=2.35$, the global absorption signal is still weaker compared with the NR case at $z\lesssim 17$, and the power is lower at $z\sim 21-17$, which can be attributed to accelerated IGM ionization by CHE. During EoH and EoR ($z\lesssim 16$), both CH and NR models are not sensitive to $\alpha$, and their differences are consistent with the trends discussed above for $f_{\star,\rm III}=0.01$ and $\alpha=1$. %A noticeable feature is that the power during EoR ($z\lesssim 11$) is slightly higher for $\alpha=2.35$ compared with $\alpha\lesssim 1$ in the NR case, but lower in the CH case. This arises from the different relations between $\epsilon_{\rm b}^{\rm ion}$ and $\alpha$ in the two stellar evolution models, which reflects the difference in the stellar mass dependence of production efficiency of ionizing photons (Fig.~\ref{fig:nphoton}). 
%Approaching the end of EoH and during EoR ($z\lesssim 15$),  

Fig.~\ref{fig:21cm_trec} shows that $t_{\rm rec}$ mainly affects the timings of the three epochs, as it regulates the formation history of Pop~II stars \citep{Magg2022tr}. For both CH and NR stars, the peak global absorption signal is shifted from $z\sim 17$ to $z\sim 15$ when $t_{\rm rec}$ is increased from 10~Myr to 100~Myr, delaying the transition from EoC to EoH. Similar shifts occur in the evolution of $\Delta^{2}(k=0.24\rm\ cMpc^{-1})$. The lowest point of global signal remains around $\langle T_{21}\rangle\sim -120$~mK for CH stars, while it drops from $\langle T_{21}\rangle\sim -130$~mK with $t_{\rm rec}\lesssim 30$~Myr to $\langle T_{21}\rangle\sim -135$~mK with $t_{\rm rec}=100$~Myr in the NR models due to the delay of X-ray heating. This deepening does not occur in the CH case because the delay of heating is counter balanced by the acceleration of IGM ionization. 

% plataeu --> X-ray heating is faster than ionization

In general, the effects of CHE on the 21-cm signal are weaker than those of varying $f_{\star,\rm III}$ but comparable to the effects of $\alpha$ and $t_{\rm rec}$ for $f_{\star,\rm III}\lesssim 0.003$. Stronger effects of CHE can be achieved through IGM ionization given higher $f_{\star,\rm III}$, but such cases will be in tension with the \citet{PlanckCollaboration2020} measurement of $\tau_0$ given a significant contribution ($\tau_{\rm II}\gtrsim 0.04$) of Pop~II stars to reionization (see Sec.~\ref{sec:ion}). However, it is shown in Fig.~\ref{fig:21cm_sfe} that when different values of $f_{\star,\rm III}$ are chosen to reproduce the same value of $\tau_0$ within the constraints from \citet[][]{PlanckCollaboration2020} for CH and NR Pop~III stars, the two cases show significant differences in the 21-cm signal. On the other hand, $\alpha$ and $t_{\rm rec}$ do not show this feature because they have much weaker effects on the IGM ionization history (Appendix~\ref{apdx:reion}). This implies that it is promising to jointly constrain $f_{\star,\rm III}$ and the fraction of CH stars in Pop~III by combining observations of the 21-cm signal and ionization history, although existing observations are not sensitive enough to provide meaningful results. The power spectra predicted by our simulations remains below the upper limits placed by \citet{HERACollaboration2023} by at least 2 orders of magnitude. 
We defer a detailed evaluation of the feasibility of such joint constraints considering the improved sensitivities of upcoming observations to future work \citep[for an example of such analysis see][]{Munoz2022}. 

%  which heat and ionize the intergalactic medium (IGM).
%The inclusion of CHE results in a stronger and earlier coupling of the spin temperature to the gas kinetic temperature, Tₖ, due to enhanced Lyman-α radiation. This leads to a deeper absorption trough in ΔT₂₁ compared to the NR case. For the fiducial model (f⭐,III = 0.003, α = 1), the minimum ΔT₂₁ reaches ∼−150 mK at z ∼ 17, which is ∼20 deeper and occurs ∼2 redshift units earlier than in the NR scenario. Increasing the SFE further deepens the trough and shifts it to higher redshifts, emphasizing the sensitivity of the 21-cm signal to early star formation.

% The dotted (solid) curves show the results for the NR (CH) Pop~III stellar evolution models, where the Pop~III SFE is varied in the range $f_{\star,\rm III}=10^{-4}-0.01$ with fixed $\alpha=1$ and $t_{\rm rec}=30$~Myr. Here, the results for higher $f_{\star,\rm III}$ are denoted by lighter colours. In the left panel, the measurement of $\tau_{0}\equiv\tau(z\rightarrow 1100)$ by Planck $\tau_{\rm obs}=0.0544\pm0.0073$ \citep{PlanckCollaboration2020} is shown with the dashed line and shaded regions for $1-3\sigma$ errors.

\section{Summary and discussion}\label{sec:dis}
%In this study, we investigate the impact of chemically homogeneous evolution (CHE) in metal-free Population III (Pop III) stars on the 21-cm signal and cosmic reionization. 
We use detailed stellar atmosphere models (Sec.~\ref{sec:spec}) to investigate how chemically homogeneous evolution (CHE), driven by rapid rotation, affects the UV spectra of metal-free Population III (Pop III) stars. These spectra are incorporated in state-of-the-art semi-numerical simulations (Sec.~\ref{sec:sim}) to explore the impact of Pop~III CHE on early star/galaxy formation and evolution of the intergalactic medium (IGM) probed by the 21-cm signal and reionization history. Different assumptions are considered for the formation efficiency (SFE) and initial mass function (IMF) of Pop~III stars, as well as the transition from Pop~III to metal-enriched Population~II (Pop~II) star formation. %to explore their interplay with CHE.  %Our methodology involved calculating the UV spectra of CHE Pop III stars and comparing them to non-rotating (NR) counterparts, incorporating these spectra into simulations of Cosmic Dawn to predict the 21-cm signal and reionization history.

Our results demonstrate that CHE significantly boosts the emission efficiency $\epsilon_{\rm b,III}^{\rm ion}$ of ionizing photons from Pop III stars (by a factor of $\sim 2-6$ in the stellar mass range $M_\star\in [9,300]\ \rm M_\odot$), leading to stronger constraints on Pop~III star formation from observations of reionization. In particular, we derive a general relation between the Thomson scattering optical depth $\tau_0$ with $\epsilon_{\rm b,III}^{\rm ion}$, the average escape fraction of ionizing photons $f_{\rm esc,III}$, the average number of recombinations experienced per baryon before remaining ionized $N_{\rm rec}$, and the stellar mass density $\rho_{\star,\rm III}$ of Pop~III stars ever formed \citep[see also][]{Visbal2015,Inayoshi2016}. We find that the $\rho_{\star,\rm III}$ allowed/required to explain the measurement of $\tau_0$ by \citet{PlanckCollaboration2020} is reduced by a factor of $\sim 2$ under CHE, given a log-flat top-heavy IMF expected from small-scale simulations of primordial star formation. Meanwhile, the emission efficiency $\epsilon_{\rm b,III}^{\rm Ly}$ of Lyman-band photons ($10.2-13.6$~eV) is slightly reduced by CHE. The lower $\rho_{\star,\rm III}$ and $\epsilon_{\rm b,III}^{\rm Ly}$ lead to shallower 21-cm global absorption signals, with reductions of up to $\sim 15$~mK (11\%) in peak absorption at $z\sim 17$, and the suppression of large-scale ($k\sim 0.2\ \rm cMpc^{-1}$) power at high redshifts ($z\gtrsim 25$) by a factor of a few. These effects are modulated by variations of Pop~III star formation parameters, which generally have comparable effects on the 21-cm signal and reionization as CHE, highlighting the interplay between stellar evolution (e.g., rotation and mixing leading to CHE) and star/galaxy formation (e.g., SFE, IMF, and transition from Pop~III to Pop~II star formation) in early cosmic evolution. Our results are sensitive to the escape fraction of Pop~III ionizing photons $f_{\rm esc, III}$, the recombination number $N_{\rm rec}$, and the contribution of Pop~II stars to reionization $\tau_{\rm II}$. The effects of CHE can be stronger when $f_{\rm esc, III}/(1+N_{\rm rec})$ is larger or $\tau_{\rm II}$ is smaller.

% , which complements previous studies that only consider the latter aspect or treat them as independent free parameters
Focusing on the UV emission from Pop~III stars during main sequence (MS), our work is a step towards the challenging yet promising goal of jointly constraining stellar evolution and star/galaxy formation during Cosmic Dawn from observations of the 21-cm signal and reionization. 
The main caveat is our ignorance of X-ray binaries (XRBs) from Pop~III stars, which could completely change the picture \citep{Gessey-Jones2025} considering the potentially strong dependence of their X-ray outputs on the IMF and binary statistics \citep[e.g.,][]{Fragos2013bps,Ryu2016,Liu2021binary,Sartorio2023,Liu2024}. Previous studies have shown that CHE has strong effects on binary stellar evolution \citep[]{deMink2016,Mandel2016,Marchant2017,Marchant2023,duBuisson2020,Riley2021,Qin2023,Dall'Amico2025,Vigna-Gomez2025}. However, they mostly focus on compact object mergers and suffer from large uncertainties in binary population synthesis models, especially in the metal-poor regime with sparse observational constraints and discrepant simulation results\footnote{Our simple assumption of no Pop~III XRBs is motivated by the rareness of close binaries of Pop~III stars in recent simulations of Pop~III star clusters \citep[e.g.,][]{Sugimura2020,Sugimura2023,Liu2021binary,Park2022,Park2024}. However, previous simulations with different configurations of Pop~III clusters have produced large fractions of close binaries \citep[e.g.,][]{Ryu2016}.}, such that the impact of CHE on high-$z$ XRBs is poorly understood. 
Another limitation is that our simulations only explore a limited part of the vast parameter space of semi-numerical simulations of Cosmic Dawn, focusing on Pop~III star formation. The other processes, such as escape of ionizing photons into the IGM, Pop~II star formation and X-ray heating, are modelled by constant parameters with values chosen from low-redshift observations or calibrated to reproduce the observed cosmic star formation rate density at $z\sim 6-8$. Improving the theoretical models and their calibrations in these aspects is an ongoing effort of the community \citep[e.g.,][Dasgupta et al. in prep.]{Munoz2022,Feathers2024,Dhandha2025,ZviKatz2025}, which is crucial for better understanding the role of CHE in follow-up studies.

Moreover, the broader implications of CHE extend beyond the 21-cm signal and reionization. As discussed in \citet{Liu2024che}, CHE can significantly enhance the rest-frame UV luminosities of high-$z$ star-forming galaxies, potentially explaining the overabundance of UV-bright galaxies observed by the JWST at $z\gtrsim 10$ \citep[e.g.,][]{Donnan2023,Donnan2024,Finkelstein2023,Finkelstein2024,Harikane2023,Adams2024} without invoking drastic changes to the `standard' model of galaxy formation. Moreover, CHE also regulates the post-MS and binary stellar evolution of massive stars, affecting the properties of Wolf-Rayet stars, supernovae, gamma-ray bursts, X-ray binaries, and binary compact object mergers \citep[e.g.,][]{Eldridge2012,Yoon2012,Szecsi2015,Mandel2016,deMink2016, Marchant2017,Kubatova2019, duBuisson2020,Riley2021,Qin2023,Umeda2024,Boco2025,Dall'Amico2025,Vigna-Gomez2025}. 
Finally, CHE and fast rotation in general have profound implications for the chemical enrichment of the early Universe with their unique nucleosynthetic signatures \citep[e.g.,][]{Chiappini2006,Chiappini2011,Chiappini2013,Maeder2015,Choplin2017,Choplin2019,Liu2021wind,Jeena2023,Tsiatsiou2024,Nandal2024}, which can explain the peculiar chemical abundance patterns of high-$z$ galaxies \citep{Bunker2023,D'Eugenio2023,Cameron2023,Senchyna2023,Ji2024,Schaerer2024,Topping2024} and extremely metal-poor stars in the local Universe \citep[e.g.,][]{Yoon2016,Yoon2018,Hansen2019,Dietz2021,Zepeda2022}. 

In conclusion, our findings underscore the importance of considering CHE and stellar rotation in general in models of early star formation and galaxy evolution. The enhanced UV and ionizing radiation from CHE stars as well as nucleosynthesis regulated by rotation and mixing not only impact the 21-cm signal and reionization but may also provide viable explanations for the observed properties of high-$z$ galaxies, energetic transients, and nearby extremely metal-poor stars. Future studies should further systematically and self-consistently explore the effects of CHE on various physical processes and observables to pave the way for joint observational constraints on the physics of stellar evolution and star/galaxy formation.

%Observations of the 21-cm signal and IGM ionization history in the next decades \citep[e.g.,][]{Koopmans2015} will bring us more data to eventually break the degeneracy between the physics of stellar evolution (e.g., CHE or not) and star/galaxy formation (e.g., SFE, IMF, and transition from Pop~III to Pop~II star formation). 

\section*{Acknowledgments}
The authors thank Furen Deng for useful discussions on \textsc{21cmSPACE}. 
BL gratefully acknowledges the funding of the Royal Society University Research Fellowship and the Deutsche Forschungsgemeinschaft (DFG, German Research Foundation) under Germany's Excellence Strategy EXC 2181/1 - 390900948 (the Heidelberg STRUCTURES Excellence Cluster). TGJ acknowledges the support of the Science and Technology Facilities Council (STFC) through the grant number ST/V506606/1. JD acknowledges support from the Boustany Foundation and Cambridge Commonwealth Trust in the form of an Isaac Newton Studentship. YS and GM have received funding from the European Research Council (ERC) under the European Union's Horizon 2020 research and innovation programme (grant agreement No.~833925, project STAREX). %This work excessively used the public packages \texttt{numpy} \citep{vanderWalt2011}, \texttt{matplotlib} \citep{Hunter2007}, and \texttt{scipy} \citep{2020SciPy-NMeth}. The authors wish to express their gratitude to the developers of these packages and to those who maintain them.

%%%%%%%%%%%%%%%%%%%%%%%%%%%%%%%%%%%%%%%%%%%%%%%%%%
\section*{Data Availability}
The full spectra $L_{\nu}(t)$, production rates of UV photons in 5 bands, and underlying stellar properties of CH Pop~III stars during MS are available at \href{https://doi.org/10.5281/zenodo.15111273}{10.5281/zenodo.15111273}, alongside with the MS lifetime-average production rates of UV photons $M_\star\epsilon_{\rm b}/t_{\rm MS}$ in 7 bands for the NR Pop~III stars modelled by \citet{Gessey-Jones2022}. For the latter (NR Pop~III stars), the Lyman-band lifetime-integrated spectra $\epsilon(\nu;M_\star)$ are available at \href{https://zenodo.org/records/5553052}{10.5281/zenodo.5553052}.   
%The code and data underlying this paper will be shared on reasonable request to the corresponding author.
 
%The inclusion of a Data Availability Statement is a requirement for articles published in MNRAS. Data Availability Statements provide a standardised format for readers to understand the availability of data underlying the research results described in the article. The statement may refer to original data generated in the course of the study or to third-party data analysed in the article. The statement should describe and provide means of access, where possible, by linking to the data or providing the required accession numbers for the relevant databases or DOIs.



\bibliographystyle{mnras}
\bibliography{ref} % if your bibtex file is called example.bib

%\begin{thebibliography}{99}
%\end{thebibliography}

%%%%%%%%%%%%%%%%%%%%%%%%%%%%%%%%%%%%%%%%%%%%%%%%%%

%%%%%%%%%%%%%%%%% APPENDICES %%%%%%%%%%%%%%%%%%%%%

\appendix
\section{Extrapolation scheme for the spectra of massive CHE stars}
\label{apdx:ext}

\begin{figure}
    \centering
    \includegraphics[width=\columnwidth]{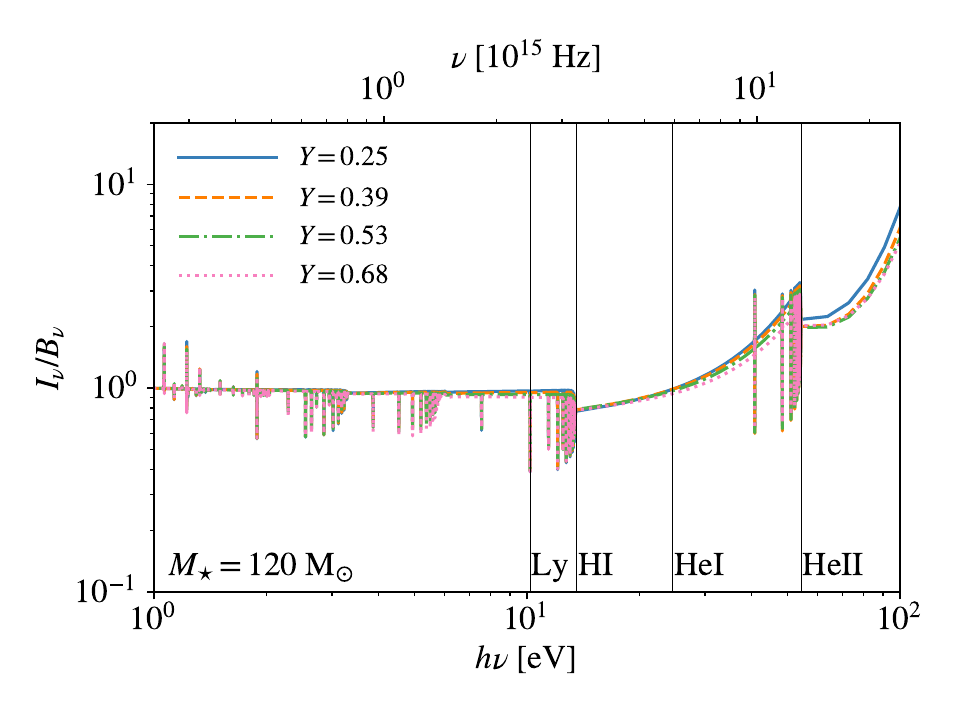}
    \vspace{-20pt}
    \caption{Deviation of spectral shape from black-body as a function of $Y$ for the $M_{\star}=120\ \rm M_\odot$ CHE model, in terms of the ratio of the intensity predicted by \sacode $I_{\nu}$ and the black-body intensity $B_{\nu}$, normalized to have $I_{\nu}/B_{\nu}=1$ at $h_{\rm P}\nu=1\ \rm eV$. The vertical lines label the characteristic energies for Lyman-band (Ly, $h_{\rm P}\nu\sim 10.2-13.6$~eV), hydrogen (HI, $h_{\rm P}\nu>13.6$~eV), helium first (HeI, $h_{\rm P}\nu>24.6$~eV) and second (HeII, $h_{\rm P}\nu>54.4$~eV) ionizing photons. The weak evolution of the shape of $I_{\nu}/B_{\nu}$ with $Y$ (and time) across a broad photon energy range ($1-100$~eV) allows us to derive the late-stage spectra of massive CH stars (that the stellar atmosphere code \textsc{tlusty} cannot handle) by simple extrapolation (see Eq.~\ref{eq:extra} and the text below).}
    \label{fig:ratio}
\end{figure}

We take the spectrum (i.e., specific luminosity) $L_{\nu}(t_{\rm ref})$ at the last time step $t_{\rm ref}$ with converged \sacode results as the reference and calculate the spectrum $L_{\nu}(t)$ at a later time $t$ with
\begin{align}
    L_{\nu}(t)=L_{\nu}(t_{\rm ref})A(t)\left[B_{\nu}(T_{\rm eff}(t))/B_{\nu}(T_{\rm eff}(t_{\rm ref}))\right]^{\beta}\ ,\label{eq:extra}
\end{align}
where $B_{\nu}(T_{\rm eff})$ is the black-body spectrum for the temperature $T_{\rm eff}$, $A(t)$ is a normalization factor set by the bolometric luminosity $L(t)$ with $\int_{0}^{\infty} L_{\nu}(t)d\nu=L(t)$, and the power-law index $\beta$ is a free parameter that describes how strongly the spectral shape evolves. Since the bolometric luminosity $L(t)$ and effective temperature $T_{\rm eff}(t)$ as functions of time throughout MS are already known from the analytical CHE model, our only task is to determine $\beta$. Interestingly, we find that the deviation of \textit{spectral shape} from black-body is almost constant for atmospheres of CH stars with converged \sacode results, as shown in Fig.~\ref{fig:ratio} for the $M_{\star}=120\ \rm M_\odot$ case as an example. To be specific, we find $I_{\nu}(t)/B_{\nu}(T_{\rm eff}(t))\approx C(t)f_{\nu}$, where $I_{\nu}=L_{\nu}/(4\pi^{2}R_{\star}^{2})$ is the stellar surface intensity given the stellar radius $R_{\star}$, $C(t)$ is a normalization factor, and $f_{\nu}$ does not evolve with time. This implies that $L_{\nu}(t)/L_{\nu}(t_{\rm ref})\approx A(t)B_{\nu}(T_{\rm eff}(t))/B_{\nu}(T_{\rm eff}(t_{\rm ref}))$, i.e., $\beta=1$, which we take as the fiducial case. 

To evaluate the uncertainty introduced by this extrapolation, we further consider two alternative cases with $\beta=0$ (no evolution) and $\beta=2$ (enhanced evolution). We find by numerical experiments that our final results are insensitive to the choice of $\beta$ for $\beta\in [0,2]$ (Sec.~\ref{sec:res}). This partially due to the fact that $T_{\rm eff}$ only varies moderately ($\lesssim 10$\%) across the extrapolation for each star. Below, we only show the results for the fiducial case $\beta=1$. Varying $\beta$ in the range of $[0,2]$ only changes $\epsilon_{\rm b}(M_\star)$ for any band considered in Sec.~\ref{sec:spec} by no more than $\sim 10$\% and does not affect our conclusions. 

\section{Comparison with the results from black-body spectra}
\label{apdx:bb}

\begin{figure}
    \centering
    \includegraphics[width=\columnwidth]{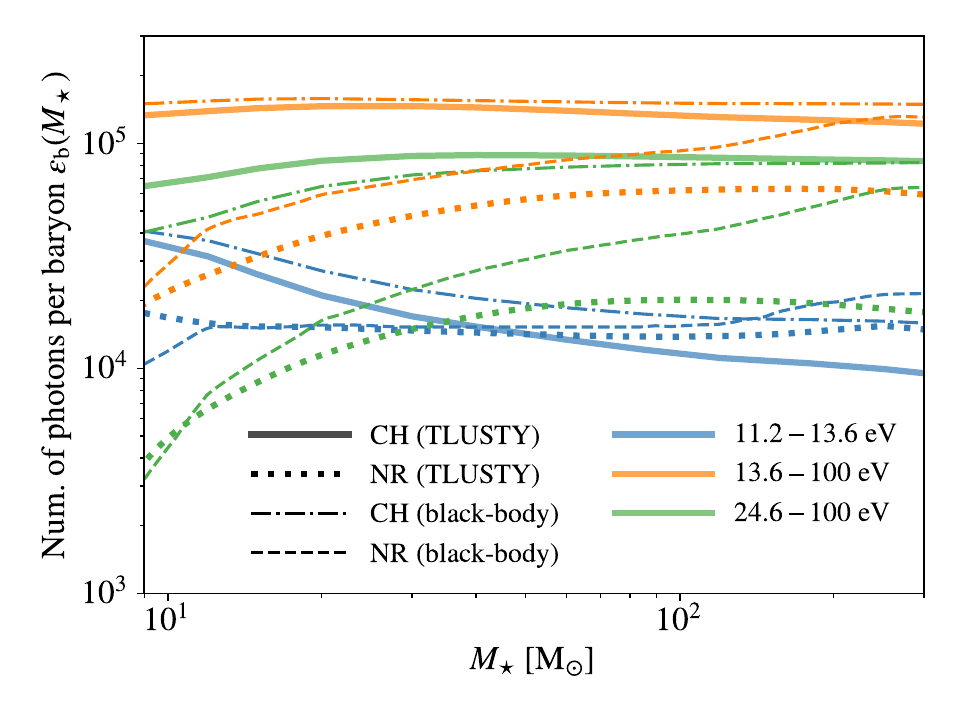}
    \vspace{-20pt}
    \caption{Similar to Fig.~\ref{fig:nphoton} but further showing the results from black-body spectra as the thin dash-dotted and dashed curves for CH and NR stars. For conciseness, only 3 bands are considered here: LW ($11.2-13.6$~eV, blue), ionizing radiation for hydrogen ($13.6-100$~eV, orange) and helium ($24.6-100$~eV, green). The black-body results for NR Pop~III stars are taken from \citet[see their fig.~2]{Sibony2022} based on the NR tracks evolved by the \textsc{genec} code with no mass loss from \citet{Murphy2021}.}
    \label{fig:nphoton_full}
\end{figure}

In general, compared with the spectrum computed by \sacode, the black-body spectrum, i.e., $L_{\nu}=4\pi^{2}R_{\star}^{2}B_{\nu}(T_{\rm eff})$, is slightly softer for a CH star but harder for an NR star. As a result, the enhancement in hydrogen ionizing photons by CHE is slightly weaker, while the enhancement in helium ionizing photons is significantly weaker, as shown in Fig.~\ref{fig:nphoton_full}. Besides, the emission efficiency of Lyman-band photons $\epsilon_{\rm b}^{\rm Ly}(M_\star)$ is increased by CHE for $M_{\star}\lesssim 200\ \rm M_\odot$ by up to a factor of $\sim 5$ under black-body approximation, while the effect of CHE is more complex and weaker in the \sacode results. This highlights the importance of detailed stellar atmosphere modelling for Lyman-band photons. 

\section{Impact of the Pop~III IMF slope and Pop~II recovery time on reionization}
\label{apdx:reion}

\begin{figure*}
    \centering
    \includegraphics[width=1\linewidth]{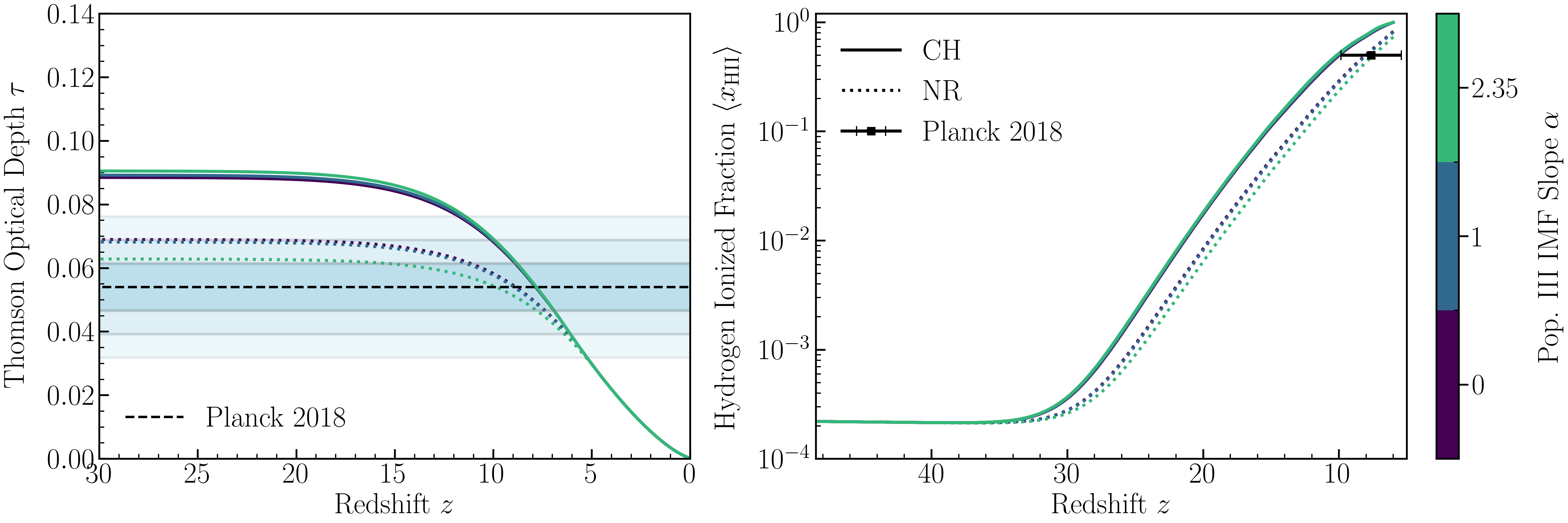}
    %{21cmSPACE_results/reion/reion_alpha.png}
    \vspace{-15pt}
    \caption{Ionization history. Same as Fig.~\ref{fig:reion_sfe} but varying the Pop~III IMF slope $\alpha$ (0, 1, and 2.35) with fixed SFE $f_{\star,\rm III}=0.003$ and recovery time $t_{\rm rec}=30$~Myr.}
    \label{fig:reion_alpha}
\end{figure*}

\begin{figure*}
    \centering
    \includegraphics[width=1\linewidth]{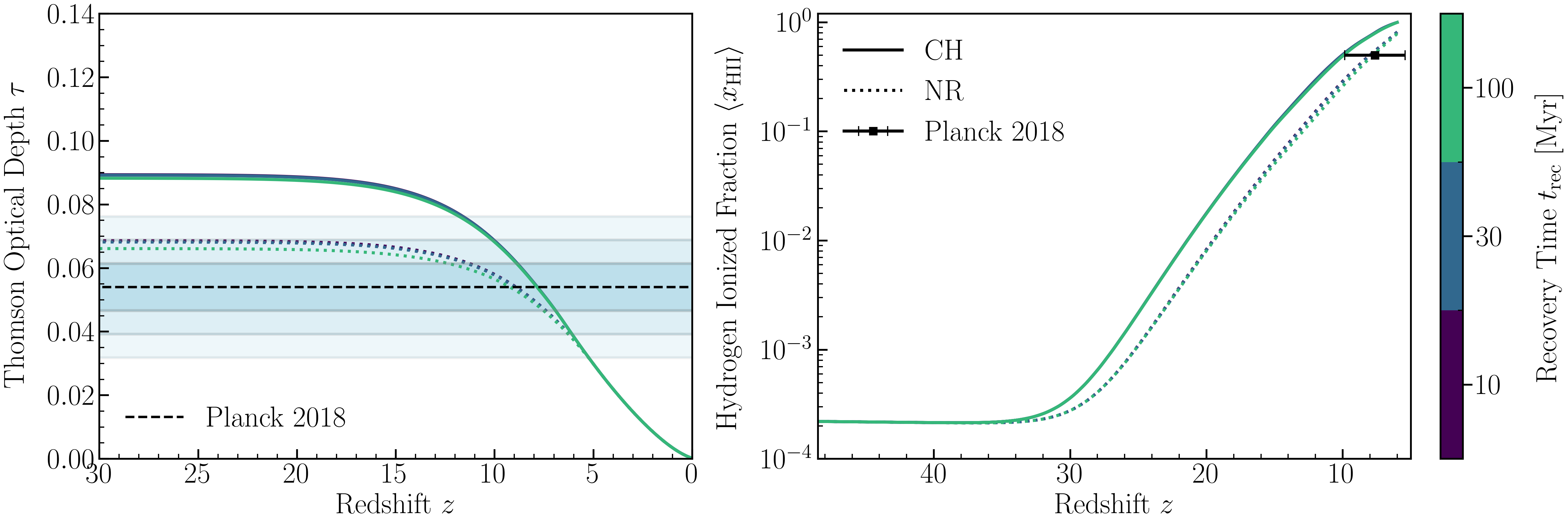}
    %{21cmSPACE_results/reion/reion_trec.png}
    \vspace{-15pt}
    \caption{Ionization history. Same as Fig.~\ref{fig:reion_sfe} but varying the recovery time $t_{\rm rec}$ (10, 30, and 100~Myr) with fixed Pop~III SFE $f_{\star,\rm III}=0.003$ and IMF slope $\alpha=1$.}
    \label{fig:reion_trec}
\end{figure*}

Fig.~\ref{fig:reion_alpha} shows the effects of the Pop~III IMF slope $\alpha$ on IGM ionization history, where three values $\alpha=0$ (top-heavy), 1 (fiducial:log-flat), and $2.35$ (bottom-heavy) are considered, under the fiducial SFE $f_{\star,\rm III}=0.003$ and recovery time $t_{\rm rec}=30$~Myr. $\alpha$ makes very small differences ($\Delta t_0\lesssim 0.005$) but in different directions for the CH and NR models. With larger $\alpha$ IGM ionization is delayed in the NR case, especially for $\alpha\gtrsim 1$, while it is slightly accelerated for the CH models. As shown in Fig.~\ref{fig:nphoton}, the production efficiency $\epsilon_{\rm b}^{\rm ion}(M_\star)$ of ionizing photons increases (decreases slightly) for more massive NR (CH) Pop~III stars, leading to different trends in the relation between the IMF-averaged efficiency $\epsilon_{\rm b,III}^{\rm ion}$ and $\alpha$ (Table~\ref{tab:epsilon}). The effect of $\alpha$ on $\tau$ is weaker for CH stars compared with the NR case because given $f_{\star,\rm III}=0.003$, IGM ionization is rapid with CH stars, such that the relation between $\tau$ and $\epsilon_{\rm b,III}^{\rm ion}$ becomes sub-linear due to saturation and self-regulation of Pop~III star formation by photo-heating feedback. %On the other hand, the production efficiency of LW photons decreases with initial stellar mass in the CH case, while it remains almost constant in the NR case. Therefore, larger $\alpha$ enhances (suppresses) radiative feedback from CH (NR) stars. 

Fig.~\ref{fig:reion_trec} shows the results for different recovery times $t_{\rm rec}=10$, 30, and 100~Myr given the fiducial Pop~III SFE $f_{\star,\rm III}=0.003$ and IMF slope $\alpha=1$. The impact of $t_{\rm rec}$ is even smaller with the same trend in the CH and NR cases: Ionization is delayed by larger $t_{\rm rec}$. The reason is that increasing $t_{\rm rec}$ only suppresses Pop~II star formation at $z\gtrsim 10$ (see Fig.~\ref{fig:sfrd_trec}) when ionization is still dominated by Pop~III stars. In conclusion, $\alpha$ and $t_{\rm rec}$ only have minor effects ($\lesssim 10$\%) on the ionization history compared with those of SFE and stellar evolution (Sec.~\ref{sec:ion}). However, they play more important roles in shaping the 21-cm signal (Sec.~\ref{sec:21cm}).

\section{Power spectra of the 21-cm signal VS wave-number at redshifts 8 and 10}
\label{apdx:21cm}

\begin{figure*}
    \centering
    \includegraphics[width=1\linewidth]{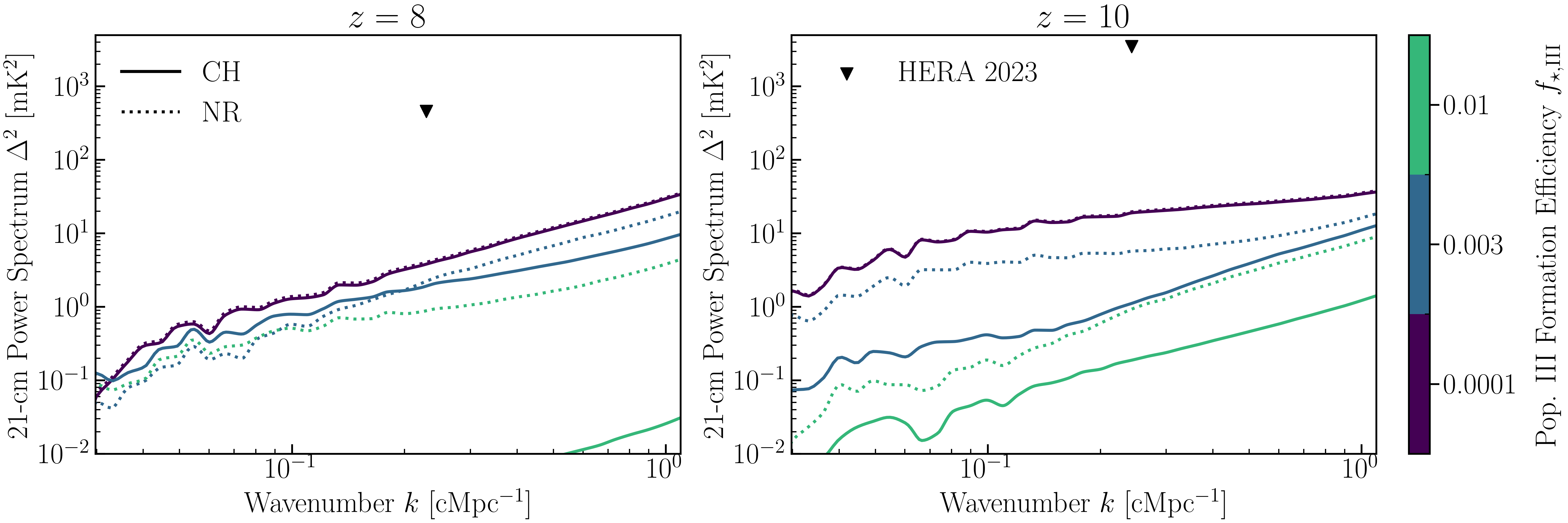}
    %{21cmSPACE_results/21cm/delta2k_HERA_fstarIII_crit.png}
    \vspace{-15pt}
    \caption{21-cm power spectrum for $z=8$ (left) and $z=10$ (right) with varying $f_{\star,\rm III}\sim 10^{-4}-0.01$ and fixed $\alpha=1$ and $t_{\rm rec}=30$~Myr. As in Fig.~\ref{fig:21cm_sfe}, the dotted (solid) curves show the results for the NR (CH) Pop~III stellar evolution models, and the results for higher $f_{\star,\rm III}$ are denoted by lighter colours. $2\sigma$ upper limits observed by \citet{HERACollaboration2023} are shown by the triangles.}
    \label{fig:ps_sfe}
\end{figure*}

\begin{figure*}
    \centering
    \includegraphics[width=1\linewidth]{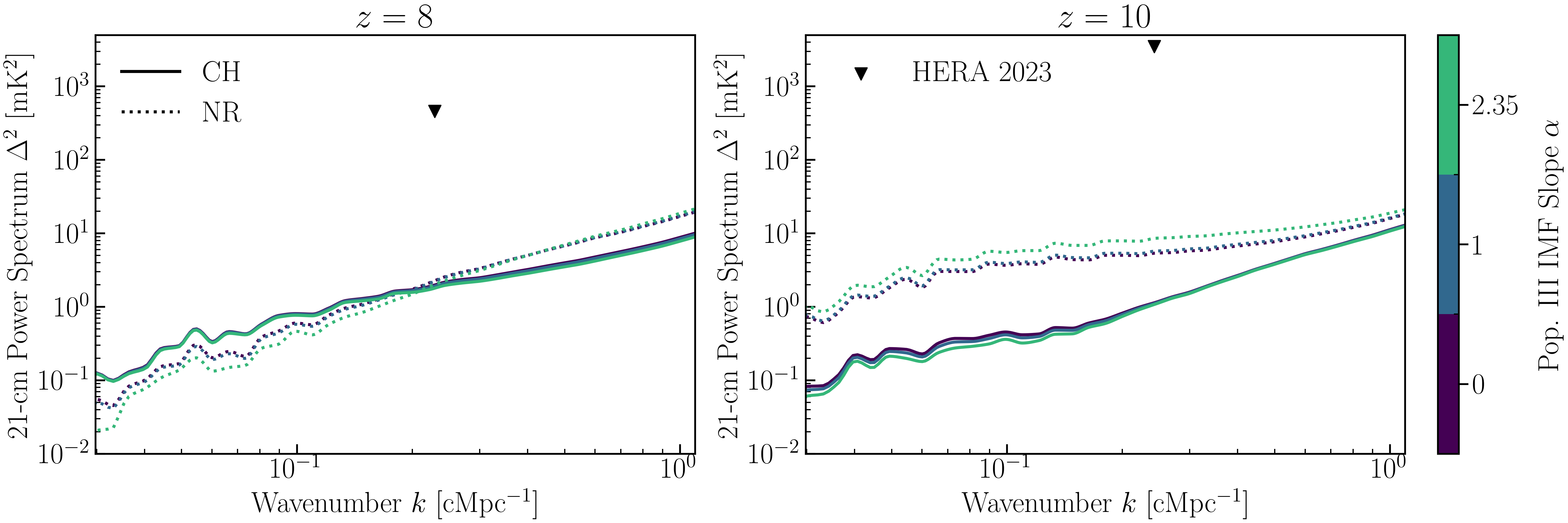}%{21cmSPACE_results/21cm/delta2k_HERA_alpha.png}
    \vspace{-5pt}
    \caption{21-cm power spectrum. Same as Fig.~\ref{fig:ps_sfe} but varying the Pop~III IMF slope $\alpha$ (0, 1, and 2.35) with fixed SFE $f_{\star,\rm III}=0.003$ and recovery time $t_{\rm rec}=30$~Myr.}
    \label{fig:ps_alpha}
\end{figure*}

\begin{figure*}
    \centering
    \includegraphics[width=1\linewidth]{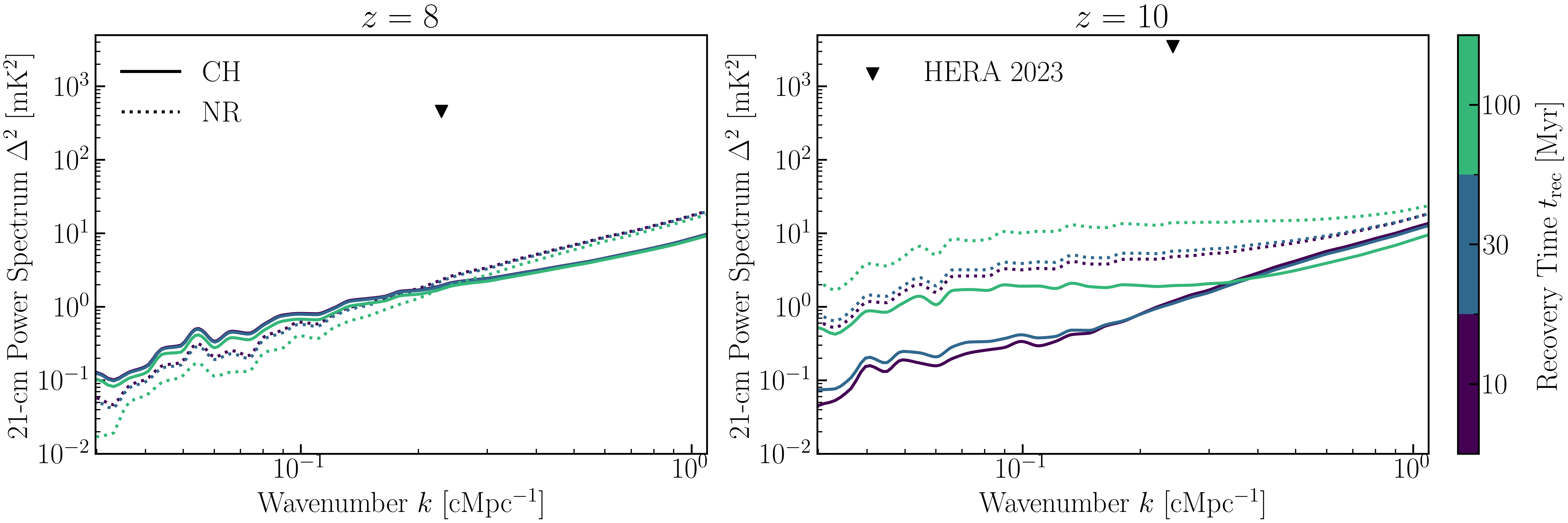}
    %{21cmSPACE_results/21cm/delta2k_HERA_trec.png}
    \vspace{-15pt}
    \caption{21-cm power spectrum. Same as Fig.~\ref{fig:ps_sfe} but varying the recovery time $t_{\rm rec}$ (10, 30, and 100~Myr) with fixed Pop~III SFE $f_{\star,\rm III}=0.003$ and IMF slope $\alpha=1$.}
    \label{fig:ps_trec}
\end{figure*}

Figs.~\ref{fig:ps_sfe}, \ref{fig:ps_alpha}, and \ref{fig:ps_trec} show the power spectra of the 21-cm signal for $k\sim 0.03-1\ h\ \rm cMpc^{-1}$ at two redshifts: $z=8$ (left panel) and 10 (right panel), where $f_{\star,\rm III}$, $\alpha$, and $t_{\rm rec}$ are varied, receptively, while keeping the other parameters fixed to their fiducial values. Here, the general spectral shape mainly reflects the transition from EoH to EoR. %, while the wiggles at large scales ($k\lesssim 0.2\ h\ \rm cMpc^{-1}$) show the imprints of acoustic oscillations from the streaming motion between baryons and dark matter, which regulates star formation and the relevant radiation fields that shape the 21-cm signal \citep[e.g.,][]{Tseliakhovich2010,Barkana2011,Naoz2011,Naoz2013,Fialkov2012,Munoz2022}. 
%that become imprinted on to the SFRD (through the feedback described in Section 2), and thus on the 21-cm signal. \citep{Munoz2022}
Before the transition, deep in EoH, the power spectrum is relatively flat with less than a factor of 10 variations across $k\sim 0.03-1\ h\ \rm cMpc^{-1}$. During the transition, the large-scale power decrease significantly due to the negative contribution of cross correlations between neutral fraction and IGM temperature fluctuations. The spectrum becomes flatter approaching the end of transition, when the distribution of neutral/ionized regions becomes the dominant factor. Thereafter, the signal decays to zero when the IGM is fully ionized. 

All models considered here are experiencing the transition from
EoH to EoR at $z=8$, which is typically in a more advanced stage for higher $f_{\star,\rm III}$ and for CH stars compared with NR stars given $f_{\star,\rm III}\gtrsim 0.003$. The effects of $\alpha$ and $t_{\rm rec}$ are always small. The transition has not happened at $z=10$ for the models that are consistent with the observational constraints from reionization (i.e., $f_{\star,\rm III}\lesssim 0.002$ for CH star and $f_{\star,\rm III}\lesssim 0.004$ for NR stars), where the volume-averaged ionized fraction of the IGM remains below 30\% (Fig.~\ref{fig:reion_sfe}). However, the other, more extreme models show the transition feature at $z=10$, where the power is suppressed by CHE in the full range $k\sim 0.03-1\ h\ \rm cMpc^{-1}$ considered here with stronger effects at larger scales (smaller $k$). At $z=10$, the large-scale power increases with $t_{\rm rec}$, which delays X-ray heating and enhances IGM temperature fluctuations. The effects of $\alpha$ are still negligible, while $f_{\star,\rm III}$ only has a significant impact when the transition happens (for $f_{\star,\rm III}\gtrsim 0.003$).  %The effects of $\alpha$

For all models considered here, the predicted $\Delta^{2}(k)$ at $z\sim 8-10$ is not only below the upper limit placed by \citet{HERACollaboration2023} at $k\simeq 0.24\ \rm cMpc^{-1}$ by at least 2 orders of magnitude (triangles in Figs.~\ref{fig:ps_sfe}-\ref{fig:ps_trec}) but also lower than the upper limit $\Delta^{2}_{\rm LOFAR}\sim 2900-4700~\rm mK^{2}$ from the Low-Frequency Array at a larger scale $k\sim 0.05\ \rm cMpc^{-1}$ \citep{Mertens2025} by at least a factor of $\sim 1000$. 

% wiggles in the 21-cm power spectrum. These are due to the streaming velocities vcb, which have acoustic oscillations that become imprinted on to the SFRD (through the feedback described in Section 2), and thus on the 21-cm signal.

%%%%%%%%%%%%%%%%%%%%%%%%%%%%%%%%%%%%%%%%%%%%%%%%%%
% Don't change these lines
\bsp	% typesetting comment
\label{lastpage}
\end{document}